\documentclass[12pt]{article}
\usepackage{amsmath}
\usepackage{amssymb}

\renewcommand{\overline}[1]{\bar{#1}}
\makeatletter\@addtoreset{equation}{section}
\makeatother
\begin{document}
\begin{titlepage}
\begin{flushright}
TIT/HEP-597\\
August, 2009\\
\end{flushright}
\vspace{0.5cm}
\begin{center}
{\Large \bf
Instanton Calculus in Deformed $\mathcal{N} = 4$ Super Yang-Mills Theories
}
\lineskip .75em
\vskip1.0cm
{\large Katsushi Ito${}^{1}$, Hiroaki Nakajima${}^{2}$, Takuya Saka${}^{1}$ and Shin
Sasaki${}^{1}$ }
\vskip 2.5em
${}^{1}$ {\normalsize\it Department of Physics\\
Tokyo Institute of Technology\\
Tokyo, 152-8551, Japan} \vskip 1.5em
${}^{2}$ {\normalsize\it BK21 Physics Research Division and Institute of Basic Science\\
Sungkyunkwan University\\
Suwon, 440-746, Korea
}
\vskip 3.5em
\end{center}
\begin{abstract}
We study the instanton effective action of four-dimensional deformed
$\mathcal{N} = 4$ supersymmetric Yang-Mills theory in the presence of
constant, self-dual Ramond-Ramond (R-R) 3-form background in type IIB
superstring theory.
We compare the instanton effective action with the
low-energy effective action on D($-1$)-branes in
the D3-D($-1$) system in the same background.
We find that discrepancy appears at the second order of the R-R
background, which was also observed in deformed $\mathcal{N} = 2$ theory.
This discrepancy is resolved if the action of the deformed gauge theory
is improved by introducing a term with coordinate-dependent gauge coupling.
We obtain the same deformed instanton effective action
from super Yang-Mills theory
in ten-dimensional $\Omega$-background by dimensional reduction.
We also discuss another type of  R-R 3-form background
which corresponds to massive deformations of the instanton effective action.

\end{abstract}
\end{titlepage}

\baselineskip=0.7cm
\section{Introduction}
It has been known that closed string backgrounds induce non-trivial effects on
D-branes, which are useful to investigate non-perturbative effects
in gauge theories.
The low-energy effective field theories on D-branes in closed string
backgrounds are described by deformed
gauge theories.
For example, the field theory on D-branes
in the constant NS-NS B-field background
is described by noncommutative gauge theory \cite{ChHo, SeWi}.
It was shown that noncommutativity resolves the small
instanton singularity in the moduli space of instantons \cite{NeSc}.

Since superstring theory contains various Ramond-Ramond (R-R) closed
string backgrounds, it would be interesting to study supersymmetric
gauge theories deformed in R-R backgrounds.
These deformed theories are useful to investigate non-perturbative
effects in field theories and also stringy effects from the viewpoint
of field theory.
Constant self-dual
graviphoton background, for example,
plays an important role to study F-terms in
$\mathcal{N} = 1$ supersymmetric gauge theories
\cite{BeCeOoVa, AnGaNaTa, DiVa, OoVa}.
This background originates from the self-dual R-R 5-form in
type IIB superstring theory.

To obtain deformed low-energy effective theories in closed string backgrounds
by taking the zero-slope limit $\alpha' \to 0$,
we need to specify the scaling condition for the backgrounds.
For example, the constant self-dual graviphoton background
$\mathcal{F}^{(\alpha \beta)}$, where
$\alpha, \beta$ are four-dimensional spinor indices and
$(\alpha \beta)$ denotes the symmetrization with respect to $\alpha$ and $\beta$,
with fixed
$(2\pi \alpha')^{\frac{3}{2}}
\mathcal{F}^{(\alpha \beta)}$
induces
non(anti)commutativity in $\mathcal{N} = 1$ superspace
\cite{BeSe, BoGrNi}.
Supersymmetric gauge theories and deformed instantons in the
${\cal N}=1$ non(anti)commutative superspace have been studied extensively
\cite{Se, ArItOh, ArTaWa}, which are realized by D3-branes and D($-1$)-branes at the
singularity of the orbifold ${\bf C}^3/({\bf Z}_2\times{\bf Z}_2)$
in the graviphoton background \cite{BiFrPeLe}.

When we consider D3-branes at the $\mathbf{C}^2/\mathbf{Z}_2$ orbifold
singularity,
the low-energy effective theory is described
by ${\cal N}=2$ super Yang-Mills theory.
The self-dual R-R 5-form background
$\mathcal{F}^{(\alpha \beta) (IJ)}$ 
with the scaling condition $(2\pi
\alpha')^{\frac{3}{2}} \mathcal{F}^{(\alpha \beta) (IJ)} =
\mathrm{fixed}$,
where $I,J=1,2$ are $SU(2)_{R}$ R-symmetry indices,
induces
${\cal N}=2$
non(anti)commutative superspace with non-singlet deformation
\cite{ItSa, IvLeZu}.
The constant R-R 1-form field strength background $\mathcal{F}^{[\alpha \beta]
[IJ]}$, where $[\alpha \beta]$ denotes the anti-symmetrization with respect to $\alpha$ and $\beta$,
is expected to provide the
singlet deformation (see for example \cite{FeIvLeSoZu}).
The low-energy effective theory of D3-branes in the self-dual R-R 5-form
background ${\cal F}^{(\alpha\beta)(AB)}$ ($A,B=1,\cdots,4$)
on the flat ten-dimensional spacetime has been studied in \cite{ItKoSa, Im},
which would correspond to non(anti)commutative
${\cal N}=4$ super Yang-Mills theory \cite{Abbaspur:2005hk, Chu:2008qa, Saemann:2004cf}.
Here $A$ and $B$ are $SU(4)_R$ R-symmetry indices.

Recently,
low-energy effective theories on D-branes in
the constant R-R 3-form field strength background have been attracted much
attentions in study of non-perturbative effects in supersymmetric gauge theories.
There are two types of the constant R-R 3-form field strength:
$\mathcal{F}^{(\alpha \beta)[IJ]}$ and
$\mathcal{F}^{[\alpha \beta] (IJ)}$.
We call these (S,A) and (A,S)-type background, respectively.
In \cite{BiFrFuLe}, they studied
the low-energy effective action of D($-1$)-branes
in the D3-D$(-1)$ system located at the
singularity of the orbifold $\mathbf{C}^2/\mathbf{Z}_2$ in the
(S,A)-background
with the scaling condition $(2\pi \alpha')^{\frac{1}{2}} \mathcal{F}^{(\alpha\beta)[IJ]} =
\mathrm{fixed}$
and found that the
action coincides with the instanton effective action of ${\cal N}=2$
super Yang-Mills theory in the $\Omega$-background.
This deformed instanton effective action is very useful to
obtain the exact formula of the prepotential via localization technique \cite{Ne, NeOk}.

In \cite{ItNaSa1}, we studied the deformed $\mathcal{N} = 2$ and
$\mathcal{N} = 4$ super Yang-Mills theories as the low-energy
effective action of the D3-branes in the (S,A)
and (A,S)-type backgrounds with
the same scaling condition as in \cite{BiFrFuLe}.
Then it is natural to expect
that the instanton effective action derived from the
D3-D$(-1)$ system can be also calculated from the ADHM construction of
instantons \cite{AtHiDrMa} of the deformed gauge theory.
However, in \cite{ItNaSa2, proItNaSa} we found that there exists a
discrepancy between the ADHM construction of instanton effective action
in deformed ${\cal N}=2$ super Yang-Mills theory and
the D($-1$)-brane effective action in the D3-D($-1$) system at the second order in
the R-R background.
This is due to the fact that the 
(S,A)-background field gives the mass term to the moduli corresponding to the position of D($-1$)-branes,
which breaks the translational symmetry.
We showed that if we improve the action of the deformed gauge theory by
adding a term with coordinate-dependent gauge coupling constant, 
we can recover the instanton effective action
obtained from the D3-D$(-1)$ system.
Although we do not know yet string theoretical derivation of this
improvement term,
the deformed gauge theory in the (S,A)-background provides a
simple method to obtain the deformed instanton effective action from the D-brane
configuration.
Therefore it is interesting to investigate whether this kind of improvement of
the deformed action happens in other theories.

In this paper, we will generalize the ${\cal N}=2$
calculations \cite{ItNaSa2, proItNaSa}
to the case of deformed $\mathcal{N} = 4$ super Yang-Mills
theories. Since $\mathcal{N} = 4$ theories have
maximal supersymmetry, the (S,A) and (A,S)-type
backgrounds admit more general deformations and the
deformed ${\cal N}=2$ theories can be obtained by the orbifold projection.
The deformed effective action of the D$(-1)$-branes embedded in the D3-branes is
evaluated by the open string disk amplitudes with background corrections and is compared with
the ADHM instanton calculus based on the deformed D3-brane action.
As in the case of the $\mathcal{N} = 2$ theory, 
we find
a disagreement between them at the second order in the deformation
parameter.
This disagreement can be resolved by adding a
term with spacetime coordinate-dependent gauge coupling,
which we call the improvement of the action.
We then find that both approaches give the same result.
We also find that this instanton effective action is obtained
from $\mathcal{N} = 4$ gauge theory in the $\Omega$-background
defined in ten dimensions, which is a natural generalization of the
six-dimensional $\Omega$-background in $\mathcal{N} = 2$ theory \cite{Ne}.

We will also discuss the (A,S)-type deformation of the instanton effective
action.
In the case of the (S,A)-type deformation, certain bosonic ADHM moduli have mass
term, whose mass depends on the deformation parameters.
In the case of the (A,S)-type, the chiral fermions in deformed super
Yang-Mills
theory have mass term.
In fact, the (A,S)-deformed instanton effective action becomes that of
mass deformed $\mathcal{N}=4$ super Yang-Mills theories.
In the instanton effective action,
this corresponds to mass term for certain fermionic moduli.

The organization of this paper is as follows.
In the next section, we evaluate open string disk amplitudes
corresponding to the string that at least one of the end points is attached
on the D$(-1)$-branes and calculate the deformed instanton effective
action up to the second order in the deformation parameter.
In section 3, we study
the (S,A)-type deformed $\mathcal{N} = 4$ super Yang-Mills theory on the D3-branes
and the ADHM construction of
instantons.
In section 4, we  study
the $\Omega$-background deformation of $\mathcal{N}=4$ super Yang-Mills theory
and compare the results with those obtained in section 2 and 3.
In section 5,
we study the deformation in
the (A,S)-type background.
Section 6 is devoted to conclusions and discussion.
Detailed calculations of open string disk amplitudes are presented in
appendix A. A brief introduction of the ADHM construction can be found
in appendix B.

\section{(S,A)-deformed
$\mathcal{N} = 4$ instanton effective action from string amplitudes}
In this section we calculate the instanton effective action based on the low-energy effective action
of the D3-D($-1$) system deformed in the (S,A)-type background.
Firstly we summarize notations and conventions used in this paper.
We denote $\alpha,\dot{\alpha}=1,2$ spinor indices of four dimensional Euclidean spacetime
$x^m=(x^1,x^2,x^3,x^4)$, which is the worldvolume of D3-branes.
We follow the notation of \cite{WeBa}.
Euclidean sigma matrices are $\sigma_{m\alpha\dot{\alpha}}=(i\tau^1,i\tau^2,i\tau^3,1)$ and
$\bar{\sigma}_m^{\dot{\alpha}\alpha}=(-i\tau^1,-i\tau^2,-i\tau^3,1)$, where $\tau^c$ ($c=1,2,3$) are the Pauli matrices.
They satisfy
$\sigma^m\bar{\sigma}^n+\sigma^n\bar{\sigma}^m=\bar{\sigma}^m\sigma^n+\bar{\sigma}^n\bar{\sigma}^m=2\delta^{mn}$.
The Lorentz generators are $\sigma_{mn}=\frac{1}{4}(\sigma_{m}\bar{\sigma}_{n}-\sigma_{n}\bar{\sigma}_{m})$ and
$\bar{\sigma}_{mn}=\frac{1}{4}(\bar{\sigma}_{m}\sigma_{n}-\bar{\sigma}_{n}\sigma_{m})$.
The {}'t Hooft symbols $\eta^{c}_{mn}$ and $\bar{\eta}^{c}_{mn}$ are defined by
$\sigma_{mn}=\frac{i}{2}\eta^{c}_{mn} \tau^{c}$ and
$\bar{\sigma}_{mn}=\frac{i}{2}\bar{\eta}^{c}_{mn} \tau^{c}$.
We use $A=1,\ldots,4$ for $SU(4)_R$ R-symmetry indices for the rotation in six-dimensional space $x^a=(x^5,\ldots,x^{10})$, which represents
transverse directions of D3-branes.
The six-dimensional matrices $\Sigma^a$ and $\bar{\Sigma}^a$ ($a=5,\ldots,10$) are 
\begin{equation}
(\Sigma^a )_{AB}=
\left( \eta^{3}, - i\bar{\eta}^{3}, \eta^{2}, - i\bar{\eta}^{2},
\eta^{1}, i\bar{\eta}^{1} \right),
\quad (\bar{\Sigma}^{a})^{AB}=
(-\eta^3, -i\bar{\eta}^3, -\eta^2,-i\bar{\eta}^2, -\eta^1, i\bar{\eta}^1),
\label{eq:sixgamma}
\end{equation}
which satisfy $\bar{\Sigma}^a\Sigma^b+\bar{\Sigma}^b\Sigma^a=\Sigma^b\bar{\Sigma}^a+\Sigma^a\bar{\Sigma}^b=2\delta^{ab}$.
The Lorentz generators are $\Sigma^{ab}=\frac{1}{4}(\Sigma^a\bar{\Sigma}^b-\Sigma^b\bar{\Sigma}^a)$ and
$\bar{\Sigma}^{ab}=\frac{1}{4}(\bar{\Sigma}^a\Sigma^b-\bar{\Sigma}^b\Sigma^a)$.

Instantons with topological number $k$
are obtained by the ADHM construction
and are parametrized by the ADHM moduli \cite{AtHiDrMa}.
In ${\cal N}=4$ super Yang-Mills theory, the bosonic ADHM moduli are
$(a'_{\alpha\dot{\alpha}})_{ij}$ and $w_{uj\dot{\alpha}}$, where
$i,j=1,\ldots, k$ are instanton indices, $u=1,\ldots, N$ is a gauge index.
We define $(a'_m)_{ij}$ as $(a'_m)_{ij}=\frac{1}{2}\bar{\sigma}_{m}^{\dot{\alpha}\alpha} (a'_{\alpha\dot{\alpha}})_{ij}$.
In addition, we introduce bosonic auxiliary fields $\chi_a$ and $D^{c}$ ($c=1,2,3$), which are $k\times k$ complex matrices.
$D^c$ is the Lagrange multiplier for the ADHM constraints.
Fermionic ADHM moduli are $\mu_{uj}$ and $({\cal M}'{}_{\alpha}^{A})_{ij}$. 
There exist fermionic auxiliary fields $\bar{\psi}^{\dot{\alpha}}_{A}$ for the fermionic ADHM constraints.

In string theory, gauge theory instantons with instanton number $k$ are understood as
$k$ D$(-1)$-branes embedded in D3-brane worldvolume \cite{Douglas}.
The ${\cal N}=4$ ADHM moduli parameters of the instantons correspond to the zero-modes
of the open strings whose at
least one endpoint is attached to the D$(-1)$-branes.
The instanton effective action is obtained by evaluating the disk
amplitudes which contain vertex operators associated with the
ADHM moduli and
the vacuum expectation values (VEVs) of the adjoint scalars 
living on the D3-branes.
The effect of closed string backgrounds is interpreted as the
insertions of the closed string vertex operators in the disk \cite{BiFrFuLe}.

We start from reviewing the undeformed instanton effective action
in the absence
of the (S,A)-type background \cite{BiFrPeFuLeLi}.
We use the NSR formalism to calculate disk amplitudes.
The relevant vertex operators associated with the instanton moduli
are listed in Table \ref{N4ADHM} \cite{BiFrPeFuLeLi}.
\begin{table}[t]
\begin{tabular}{|l|l|l|}
\hline
Brane & Vertex Operator & Representation \\
\hline \hline
D$(-1)$/D$(-1)$ & $ \displaystyle V^{(-1)}_{a'} (y) = \frac{\pi}{\sqrt{2}} (2 \pi
\alpha')^{\frac{1}{2}} g_0 a'_{m} \psi^{m} e^{- \phi} (y) $ &
\\
& $ \displaystyle V_{\chi}^{(-1)} (y)= \frac{1}{\sqrt{2}}(2 \pi \alpha')^{\frac{1}{2}}
\chi_a \psi^a e^{- \phi } (y)$ &
\\
& $ \displaystyle V_{\mathcal{M}}^{(- 1/2)} (y)= \pi (2 \pi \alpha')^{\frac{3}{4}} g_0
\mathcal{M}^{\prime \alpha A} S_{\alpha} S_A e^{ - \frac{1}{2}
\phi} (y)$ &
$U(k)$ adjoint \\
& $ \displaystyle V_{\bar{\psi}}^{(- 1/2)} (y)= 2 (2 \pi \alpha')^{\frac{3}{4}}
\bar{\psi}_{\dot{\alpha} A} S^{\dot{\alpha}} S^A e^{- \frac{1}{2} \phi}
(y) $ &
\\
& $ \displaystyle V^{(0)}_D (y) =
2 (2 \pi \alpha') D_c \bar{\eta}^c_{mn} \psi^{n} \psi^{m} (y) $ &
\\
& $ \displaystyle V_Y^{(0)} (y) = 4\pi (2 \pi \alpha') g_0 Y_{m a} \psi^{m} \psi^a (y) $ &
\\
\hline
D3/D$(-1)$ & $\displaystyle V_{w}^{(-1)} (y)= \frac{\pi}{2} (2 \pi \alpha')^{\frac{1}{2}} g_0 w_{\dot{\alpha}}
\Delta S^{\dot{\alpha}} e^{- \phi } (y) $ &
\\
& $\displaystyle V^{(-1)}_{\bar{w}} (y) = \frac{\pi}{2} (2 \pi
\alpha')^{\frac{1}{2}} g_0 \bar{w}_{\dot{\alpha}} \overline{\Delta}
S^{\dot{\alpha}} e^{- \phi} (y) $ &
\\
& $\displaystyle V_{\mu}^{(-1/2)} (y)= \pi (2 \pi \alpha')^{\frac{3}{4}} g_0
\mu^A \Delta S_A e^{- \frac{1}{2} \phi } (y) $ &
$U(k) \times U(N) $ bi-fundamental
\\
& $\displaystyle V_{\bar{\mu}}^{(-1/2)} (y)= \pi (2 \pi \alpha')^{\frac{3}{4}} g_0
\bar{\mu}^A \overline{\Delta} S_A e^{- \frac{1}{2} \phi } (y) $ &
\\
& $ \displaystyle V_X^{(0)} (y)= 2 \sqrt{2} \pi (2 \pi \alpha') g_0 X_{\dot{\alpha} a} \Delta
S^{\dot{\alpha}} \psi^a (y)$ &
\\
& $ \displaystyle V_{\overline{X}}^{(0)} (y)=
2 \sqrt{2} \pi (2 \pi \alpha') g_0 \overline{X}_{\dot{\alpha} a}
\overline{\Delta} S^{\dot{\alpha}} \psi^a (y) $ &
\\
\hline
\end{tabular}
\caption{$\mathcal{N} = 4$ ADHM vertex operators}
\label{N4ADHM}
\end{table}
Here we denote the vertex operator for moduli
$\Phi$ in the
$q$-picture by
$V^{(q)}_{\Phi}$.
The worldsheet fermions are decomposed  into 
$\psi^m$ and
$\psi^a$.
$\phi$ is the bosonized ghost whose momentum in a vertex
operator specifies the picture number.
$\Delta$ and $\bar{\Delta}$ are twist
fields, which interchange the D3 and D$(-1)$ boundary in a disk amplitude
\cite{Twist}.
The ten-dimensional spin field is decomposed into the four-dimensional part
$S^{\alpha}, S_{\dot{\alpha}}$ and the six-dimensional part $S^A, S_A$.
More detailed explanation for the convention and notation of these
worldsheet variables can be found in \cite{ItSa, ItKoSa}.
The zero-modes from the D$(-1)$/D$(-1)$ strings belong to the adjoint representation of
$U(k)$.
Its generators $t^U$ are normalized as $\mathrm{tr}_k [t^U t^V] =
\kappa \delta^{UV}$, where the trace is taken over $k \times k$ matrices.
$g_0 = (2 \pi)^{-\frac{3}{2}} g_s^{\frac{1}{2}} \alpha^{\prime -1}$  is the coupling constant of the D$(-1)$-branes \cite{Po},
which should go to infinity in the zero-slope limit with fixed string coupling constant $g_{\mathrm{s}}$.
Note that in order to reproduce the correct field theory result, some of the moduli should
be rescaled by $g_0$ in the zero-slope limit \cite{BiFrPeFuLeLi}.
In addition to the ${\cal N}=4$ ADHM moduli fields it is convenient to introduce new auxiliary fields
$Y_{ma}, X_{\dot{\alpha} a}, \overline{X}^{\dot{\alpha}} {}_a$
to write quartic interaction terms from the cubic interactions for the auxiliary fields
as in the case of ${\cal N} = 2$ \cite{BiFrFuLe}.

From the charge conservation for the bosonized fermions in the 
correlator, it is easy to find that the non-zero amplitudes in the limit $\alpha' \to 0$
are
\begin{eqnarray}
\begin{aligned}
& \langle \! \langle V^{(0)}_{Y} V^{(-1)}_{\chi} V^{(-1)}_{a'} \rangle \!
\rangle, \
\langle \! \langle
V^{(0)}_{X} V^{(-1)}_{\chi} V^{(-1)}_{\bar{w}}
\rangle \! \rangle, \
\langle \! \langle V^{(-1)}_{w} V^{(-1)}_{\chi} V^{(0)}_{\overline{X}}
\rangle \! \rangle, \\
& \langle \! \langle V^{(-1/2)}_{\bar{\mu}} V^{(-1)}_{\chi}
V^{(-1/2)}_{\mu} \rangle \! \rangle, \
\langle \! \langle V^{(-1/2)}_{\mathcal{M}'} V^{(-1/2)}_{\mathcal{M}'}
V^{(-1)}_{\chi} \rangle \! \rangle, \
\langle \! \langle V_{\bar{\psi}}^{(-1/2)} V_{\bar{\mu}}^{(-1/2)}
V_{w}^{(-1)} \rangle \! \rangle, \\
& \langle \! \langle V_{\bar{\psi}}^{(-1/2)} V_{\bar{w}}^{(-1)}
V_{\mu}^{(-1/2)} \rangle \! \rangle, \
\langle \! \langle V_{\bar{\psi}}^{(-1/2)} V_{\mathcal{M}'}^{(-1/2)}
V_{a'}^{(-1)} \rangle \! \rangle, \
\langle \! \langle
V_D^{(0)} V_{\bar{w}}^{(-1)} V_{w}^{(-1)}
\rangle \! \rangle, \
\langle \! \langle
V_D^{(0)} V_{a'}^{(-1)} V_{a'}^{(-1)}
\rangle \! \rangle.
\end{aligned}
\label{zero-th_amplitudes}
\end{eqnarray}
Calculating all the amplitudes and taking the limit,
we find that the amplitudes in (\ref{zero-th_amplitudes}) are reproduced by the following low-energy effective action
\begin{eqnarray}
S_{\mathrm{str}}^{0}
&=& \frac{2\pi^2}{\kappa}
\mathrm{tr}_k
\left[ \frac{}{}
Y^m {}_a Y_{ma} - X_{\dot{\alpha} a} \overline{X}^{\dot{\alpha}} {}_a
\right. \nonumber \\
& & \qquad \qquad
+ 2 Y^m {}_a [\chi_a, a'_m]
- X_{\dot{\alpha} a} (\chi_a \bar{w}^{\dot{\alpha}} -
\bar{w}^{\dot{\alpha}} \phi^0_a) -
(w_{\dot{\alpha}} \chi_a - \phi^0_a w_{\dot{\alpha}})
\overline{X}^{\dot{\alpha}} {}_a
\nonumber \\
& & \qquad \qquad
+ \frac{1}{2} (\overline{\Sigma}^a)_{AB} \bar{\mu}^A
(- \mu^B \chi_a + \phi^0_a \mu^B)
- \frac{1}{2} (\overline{\Sigma}^a)_{AB} \mathcal{M}^{\prime \alpha A}
\mathcal{M}'_{\alpha} {}^B \chi_a 
\nonumber \\
& & \qquad \qquad
- i \bar{\psi}^{\dot{\alpha}} {}_A
\left(
\bar{\mu}^A w_{\dot{\alpha}}
+ \bar{w}_{\dot{\alpha}} \mu^A
+ [\mathcal{M}^{\prime \alpha A}, a'_{\alpha \dot{\alpha}}]
\right) \nonumber \\
& & \qquad \qquad \left.
-
i D^c (\tau^c)^{\dot{\alpha}} {}_{\dot{\beta}}
\left(
\bar{w}^{\dot{\beta}} w_{\dot{\alpha}}
+ \bar{a}^{\prime \dot{\beta} \alpha} a'_{\alpha \dot{\alpha}}
\right)
\right]
\label{aux_instanton}
\end{eqnarray}
where
$\phi_a^0$ are the VEVs of the adjoint fields in $\mathcal{N} = 4$
super Yang-Mills theory.
Note that the VEV-dependent amplitudes can be
calculated similarly by replacing $\chi_a \to - \phi^0_a$.

The instanton effective action (\ref{aux_instanton}) is invariant under the
supersymmetry transformations:
\begin{eqnarray}
\delta a'_{\alpha\dot{\alpha}}&=& i\bar{\xi}_{\dot{\alpha} A}{\cal M}'_{\alpha}{}^A,
\nonumber\\
\delta {\cal M}'_{\alpha}{}^A&=&
-2i \bar{\xi}^{\dot{\alpha}}_B (\Sigma^a)^{AB}Y_{m a}(\sigma^m)_{\alpha\dot{\alpha}},
\nonumber\\
  \delta Y_{ma}&=&i \bar{\xi}^{\dot{\beta}}_{C}
[\chi_b,{\cal M}'{}^{\alpha A}]
(\sigma^m)_{\alpha\dot{\beta}}
(\bar{\Sigma}^{ba})_{A}{}^C
+2 (\Sigma^a)^{AB} (\bar{\sigma}^{nm})^{\dot{\gamma}}{}_{\dot{\beta}}
\bar{\xi}^{\dot{\beta}}_B[\bar{\psi}_{\dot{\gamma} A}, a'_{n}],
\nonumber\\
\delta w_{\dot{\alpha}}&=& i\bar{\xi}_{\dot{\alpha} A}\mu^A, \nonumber \\
\delta \mu^A&=& 2i\bar{\xi}^{\dot{\alpha}}_B (\Sigma^a)^{AB}X_{\dot{\alpha} a},
\nonumber \\
\delta X_{\dot{\alpha} a}&=&
2i \bar{\xi}_{\dot{\alpha} B}(\bar{\Sigma}^{ba})_C{}^B
(\mu^C\chi_b-\phi^0_b\mu^C)
-\bar{\xi}_{\dot{\beta} A}(\Sigma^a)^{AB} (w_{\dot{\alpha}}\bar{\psi}^{\dot{\beta}}_B
-2\delta^{\dot{\beta}}_{\dot{\alpha}} w_{\dot{\gamma}}\bar{\psi}^{\dot{\gamma}}_B
),
\nonumber\\
\delta\bar{w}_{\dot{\alpha}}&=& i\bar{\xi}_{\dot{\alpha} A}\bar{\mu}^A,
\nonumber\\
\delta \bar{\mu}^A&=& 2i \bar{\xi}_{\dot{\alpha} B} (\Sigma^a)^{AB}\bar{X}^{\dot{\alpha}}_a,
\nonumber\\
\delta \bar{X}^{\dot{\alpha}}_a &=&
2i
(\bar{\Sigma}^{ba})_A{}^C
\bar{\xi}^{\dot{\alpha}}_C (
\chi_b\bar{\mu}^A
-\bar{\mu}^A\phi^0_b
)
-(\Sigma^a)^{AB} \bar{\xi}_{\dot{\beta} A}
(\bar{\psi}^{\dot{\beta}}_B\bar{w}^{\dot{\alpha}}
-2\varepsilon^{\dot{\alpha}\dot{\beta}}\bar{\psi}^{\dot{\gamma}}_B \bar{w}_{\dot{\gamma}}),
\nonumber\\
\delta \chi_a&=& (\Sigma^a)^{AB} \bar{\xi}_{\dot{\alpha} A}\bar{\psi}^{\dot{\alpha}}_B,
\nonumber\\
\delta
\bar{\psi}^{\dot{\alpha}}_A&=&
(\bar{\Sigma}^{ab})_{A}{}^B [\chi_a,\chi_b]\bar{\xi}^{\dot{\alpha}}_B
-i \vec{D}\cdot\vec{\tau}^{\dot{\alpha}}{}_{\dot{\beta}} \bar{\xi}^{\dot{\beta}}_A,
\nonumber\\
\delta \vec{D}&=& -i\vec{\tau}^{\dot{\alpha}}{}_{\dot{\beta}}(\Sigma^a)^{AB}
\bar{\xi}_{\dot{\alpha} B}[\bar{\psi}^{\dot{\beta}}_A,\chi_a].
\label{eq:n4susy3d}
\end{eqnarray}

After integrating out the auxiliary fields
$Y_{ma}, X_{\dot{\alpha}a}, \bar{X}_{\dot{\alpha}a}$
in (\ref{aux_instanton}), we obtain the low-energy effective action \cite{DoHoKh}
\begin{eqnarray}
\tilde{S}_{\mathrm{str}}^{0} &=&
\frac{2\pi^2}{\kappa} \mathrm{tr}_k
\left[ \frac{}{}
- [\chi_a, a'_m]^2
+ \left(
w_{\dot{\alpha}} \chi_a - \phi^0_a w_{\dot{\alpha}}
\right)
\left(
\chi_a \bar{w}^{\dot{\alpha}} - \bar{w}^{\dot{\alpha}} \phi^0_a
\right)
\right. \nonumber \\
& & \qquad \qquad
\left.
+ \frac{1}{2} (\overline{\Sigma}^a)_{AB} \bar{\mu}^A
(- \mu^B \chi_a + \phi^0_a \mu^B)
- \frac{1}{2} (\overline{\Sigma}^a)_{AB} \mathcal{M}^{\prime \alpha A}
\mathcal{M}'_{\alpha} {}^B \chi_a
\right] \nonumber \\
& & + S_{\mathrm{ADHM}},
\label{undeformed_inst}
\end{eqnarray}
where
\begin{eqnarray}
S_{\mathrm{ADHM}}
&=& \frac{2\pi^2}{\kappa}
\mathrm{tr}_k
\left[ \frac{}{}
- i \bar{\psi}^{\dot{\alpha}} {}_A
\left(
\bar{\mu}^A w_{\dot{\alpha}}
+ \bar{w}_{\dot{\alpha}} \mu^A
+ [\mathcal{M}^{\prime \alpha A}, a'_{\alpha \dot{\alpha}}]
\right)
\right. \nonumber \\
& & \qquad \qquad \left.
-
i D^c (\tau^c)^{\dot{\alpha}} {}_{\dot{\beta}}
\left(
\bar{w}^{\dot{\beta}} w_{\dot{\alpha}}
+ \bar{a}^{\prime \dot{\beta} \alpha} a'_{\alpha \dot{\alpha}}
\right)
\right],
\end{eqnarray}
is the terms providing the (fermionic) ADHM constraints. 
This action indeed agrees with the  instanton
effective action for $\mathcal{N} = 4$ super Yang-Mills theory based on the ADHM construction \cite{DoHoKhMa}.

Let us introduce the (S,A)-type background. 
The vertex operator corresponding to this background in the
$(-1/2, -1/2)$ picture is given as \cite{ItNaSa1}
\begin{eqnarray}
V^{(-1/2,-1/2)}_{\mathcal{F}} (z,\bar{z})=
(2 \pi \alpha') \mathcal{F}^{(\alpha \beta)[AB]} S_{\alpha} S_{A}
e^{- \frac{1}{2} \phi} (z) S_{\beta} S_{B}
e^{- \frac{1}{2} \phi} (\bar{z}),
\end{eqnarray}
where we have identified the left- and right-moving fields due to the boundary condition at $z = \bar{z}$ \cite{BiFrPeFuLeLi}.
Since the background contains two four-dimensional spin fields $S_{\alpha},
S_{\beta}$, we need to insert other vertex operators including
$S_{\alpha} S_{\beta}$ or $\psi^m$ to get non-zero results.
Otherwise the amplitude $\mathcal{A}$ behaves like
$\mathcal{A} \propto \varepsilon_{\alpha \beta} \mathcal{F}^{(\alpha
\beta) [AB]} $ and gives vanishing contribution. The candidates of such
(combinations of ) vertex operators are $V_{a'}, V_{\mathcal{M}'} V_{\mathcal{M}'}, V_{D},
V_{Y}$.
However, we should also saturate the internal $SU(4)_R$ charge.
It is impossible to do this only by
the insertions of $V_{a'}, V_{D}$.
Therefore non-zero amplitudes which contain one $V_{\mathcal{F}}^{(-1/2,-1/2)}$
must
also contain $V_{\mathcal{M}'} V_{\mathcal{M}'}$ or $V_Y$.
Considering power
counting of $\alpha'$ and $g_0$, we find
that the non-zero amplitudes after taking the zero-slope limit
are
\begin{eqnarray}
\langle \! \langle V^{(-1/2)}_{\mathcal{M}'}
V^{(-1/2)}_{\mathcal{M}'} V^{(-1/2,-1/2)}_{\mathcal{F}}
\rangle \! \rangle, \qquad
\langle \! \langle
V^{(0)}_{Y} V^{(-1)}_{a'} V^{(-1/2,-1/2)}_{\mathcal{F}}
\rangle \! \rangle.
\label{SA_amplitudes}
\end{eqnarray}
These amplitudes are evaluated in the appendix A.
After taking the zero-slope limit, the first amplitude in (\ref{SA_amplitudes}) becomes
\begin{eqnarray}
\langle \! \langle V^{(-1/2)}_{\mathcal{M}'}
V^{(-1/2)}_{\mathcal{M}'} V^{(-1/2,-1/2)}_{\mathcal{F}}
\rangle \! \rangle
= \frac{2\pi^2}{\kappa} \mathrm{tr}_k
\left[
\frac{1}{2}
(\overline{\Sigma}^a)_{AB} \mathcal{M}^{\prime A}_{\alpha}
\mathcal{M}^{\prime B}_{\beta} (2 \pi i^2) (2 \pi \alpha')^{\frac{1}{2}}
(\overline{\Sigma}^a)_{CD} \mathcal{F}^{(\alpha \beta) [CD]}
\right].
\nonumber \\
\end{eqnarray}
The second amplitude in (\ref{SA_amplitudes}) is
evaluated as
\begin{eqnarray}
\!\!\!\!  \langle \! \langle
V^{(0)}_Y V^{(0)}_{a'} V^{(-1/2,-1/2)}_{\mathcal{F}}
\rangle \! \rangle
\!=\! \frac{2\pi^2}{\kappa}
\mathrm{tr}_k \!\!
\left[
- \frac{i}{\sqrt{2}} (2 \pi i)
(\sigma^{mn})_{\alpha \beta} (\overline{\Sigma}^a)_{AB}
Y_{ma} a'_n (2 \pi \alpha')^{\frac{1}{2}} \mathcal{F}^{(\alpha \beta) [AB]}
\right].
\end{eqnarray}
These amplitudes are reproduced by the interaction terms
on the D$(-1)$-branes
induced by the (S,A)-type background,
which are given by
\begin{eqnarray}
\delta S_{\mathrm{(S,A)}}
= \frac{2\pi^2}{\kappa}
\mathrm{tr}_k
\left[
2 Y_{ma} a'_n C^{mna}
- \frac{1}{4} (\overline{\Sigma}^a)_{AB}
\mathcal{M}^{\prime A}_{\alpha} \mathcal{M}_{\beta}^{\prime B}
C^{(\alpha \beta) a}
\right],
\end{eqnarray}
where
\begin{eqnarray}
C^{mna} &=& \varepsilon_{\beta \gamma} (\sigma^{mn})_{\alpha} {}^{\gamma} (\bar{\Sigma}^a)_{AB} C^{(\alpha \beta) [AB]},
\nonumber\\
C^{(\alpha \beta) a} &=& (\bar{\Sigma}^a)_{AB} C^{(\alpha \beta) [AB]},
\nonumber\\
C^{(\alpha \beta) [AB]} &=& - 2
\pi (2 \pi \alpha')^{\frac{1}{2}} \mathcal{F}^{(\alpha \beta) [AB]}.
\label{deformation_parameter}
\end{eqnarray}
We note that $C^{mna}$
satisfies the self-dual condition
\begin{equation}
C^{mna}=\frac{1}{2}\varepsilon^{mnpq}C_{pq}{}^{a}.
\label{C_SD}
\end{equation}
The deformation term $\delta S_{\mathrm{(S,A)}}$ is added to the undeformed part
(\ref{undeformed_inst}). After integrating out the auxiliary fields $Y_{ma}, X_{\dot{\alpha}a}$ and $\bar{X}_{\dot{\alpha}a}$,
we obtain the following deformed instanton effective action
\begin{eqnarray}
\tilde{S}_{\mathrm{str}}^{C} &=& \frac{2\pi^2}{\kappa}
\mathrm{tr}_k
\left[
-
\left(
[\chi_a, a'_m] + C_{mna} a^{\prime n}
\right)^2
+
\left(
w_{\dot{\alpha}} \chi_a - \phi^0_a w_{\dot{\alpha}}
\right)
\left(
\chi_a \bar{w}^{\dot{\alpha}} - \bar{w}^{\dot{\alpha}} \phi^0_a
\right)
\right. \nonumber \\
& & \qquad \qquad
+ \frac{1}{2} (\overline{\Sigma}^a)_{AB} \bar{\mu}^A
(- \mu^B \chi_a + \phi^0_a \mu^B)
- \frac{1}{2} (\overline{\Sigma}^a)_{AB} \mathcal{M}^{\prime \alpha A}
\mathcal{M}'_{\alpha} {}^B \chi_a
\nonumber \\
& & \qquad \qquad
\left.
- \frac{1}{4} (\overline{\Sigma}^a)_{AB} C^{(\alpha \beta) a}
\mathcal{M}'_{\alpha} {}^A \mathcal{M}'_{\beta} {}^B
\right]+S_{\mathrm{ADHM}}.
\label{deformed_inst_eff_action}
\end{eqnarray}
This is a natural $\mathcal{N} = 4$ extension of
the $\mathcal{N} = 2$ deformed instanton effective action found in \cite{BiFrFuLe}.
Note that at the second order in the R-R background, there is a mass
term for the position moduli $a'_m$ implying the fact that the position
of the instantons are fixed at the origin of the D3-brane
worldvolume.

The deformed instanton effective action (\ref{deformed_inst_eff_action})
preserves half of the $\mathcal{N} = 4$ supersymmetry.
We can see that the action is invariant under the following deformed supersymmetry transformations
\begin{eqnarray}
\delta a'_{\alpha \dot{\alpha}} &=& i \bar{\xi}_{\dot{\alpha} A}
\mathcal{M}'_{\alpha} {}^A, \nonumber \\
\delta \mathcal{M}'_{\alpha} {}^A &=& 2i
\bar{\xi}^{\dot{\alpha}}_B (\Sigma^a)^{AB}
(\sigma^m)_{\alpha \dot{\alpha}} ( [\chi_a , a'_m] + a'^n C_{mn a} ), \nonumber \\
\delta w_{\dot{\alpha}} &=& i \bar{\xi}_{\dot{\alpha} A} \mu^A,
\nonumber \\
\delta \mu^A &=& - 2 i \bar{\xi}^{\dot{\alpha}}_B (\Sigma^a)^{AB}
( w_{\dot{\alpha}} \chi_a - \phi^0_a w_{\dot{\alpha}} ), \nonumber \\
\delta \bar{w}_{\dot{\alpha}} &=& i \bar{\xi}_{\dot{\alpha} A}
\bar{\mu}^A,
\nonumber \\
\delta \bar{\mu}^A &=& - 2i \bar{\xi}_{\dot{\alpha} B} (\Sigma^a)^{AB}
( \chi_a \bar{w}^{\dot{\alpha}} - \bar{w}^{\dot{\alpha}} \phi^0_a ),
\nonumber \\
\delta \chi_a &=& (\Sigma^a)^{AB} \bar{\xi}_{\dot{\alpha} A}
\bar{\psi}^{\dot{\alpha}}_B,
\nonumber \\
\delta \bar{\psi}^{\dot{\alpha}}_A &=& (\bar{\Sigma}^{ab})_A {}^B
[\chi_a, \chi_b] \bar{\xi}^{\dot{\alpha}}_B - i \vec{D} \cdot
\vec{\tau}^{\dot{\alpha}} {}_{\dot{\beta}} \bar{\xi}^{\dot{\beta}}_A,
\nonumber \\
\delta \vec{D} &=& - i \vec{\tau}^{\dot{\alpha}} {}_{\dot{\beta}}
(\Sigma^a)^{AB} \bar{\xi}_{\dot{\alpha} B} [\bar{\psi}^{\dot{\beta}}_A, \chi_a].
\label{SUSY_tranf_inst}
\end{eqnarray}
To show the invariance, the deformation parameters must satisfy the condition
\begin{eqnarray}
C^{mn} {}_a C_{npb} - C^{mn} {}_b C_{npa} = 0.
\end{eqnarray}
As we will see in section 4, this corresponds to the flatness condition for the
ten-dimensional $\Omega$-background spacetime.

\section{(S,A)-deformed $\mathcal{N} = 4$ super Yang-Mills theory}
In this section, we calculate the instanton effective action from the
(S,A)-deformed 
$\mathcal{N} = 4$ $U(N)$ super Yang-Mills theory obtained in \cite{ItNaSa1}.
Since we are interested in obtaining
instanton solutions,
we perform the Wick rotation and consider
the action in the Euclidean spacetime.
The deformed Lagrangian
takes the form
\begin{eqnarray}
\mathcal{L} = \mathcal{L}_0 + \mathcal{L}_C,
\label{N4Lag}
\end{eqnarray}
where $\mathcal{L}_C$ denotes the (S,A)-deformation terms to the $\mathcal{N}=4$ super Yang-Mills 
theory. $\mathcal{L}_0$ is the Lagrangian of 
$\mathcal{N} = 4$ $U(N)$ super
Yang-Mills theory    
\begin{eqnarray}
\mathcal{L}_0 &=& \frac{1}{\kappa}
\mathrm{Tr} \left[
\frac{1}{4} F_{mn} F^{m n} + \frac{i \theta g^2}{32 \pi^2}
F_{m n} \tilde{F}^{m n}
+ \Lambda^{\alpha A} (\sigma^{m})_{\alpha \dot{\beta}} D_{m}
\overline{\Lambda}^{\dot{\beta}}_{\ A}
+ \frac{1}{2} \left(D_{m} \varphi_a \right)^2 \right. \nonumber \\
& & \left. - \frac{1}{2}g \left( \Sigma^a \right)^{AB} \overline{\Lambda}_{\dot{\alpha}A}
[\varphi_a, \overline{\Lambda}^{\dot{\alpha}}_{\ B} ] - \frac{1}{2}g \left(
\overline{\Sigma}^a \right)_{AB} \Lambda^{\alpha A} [\varphi_a,
\Lambda_{\alpha}^B ] - \frac{1}{4}g^2 [\varphi_a, \varphi_b]^2 \right].
\label{N4SYM}
\end{eqnarray}
Here
$F_{m n} = \partial_{m} A_{n} - \partial_{n} A_{m}
+ i g [A_{m}, A_{n}]$ is the field strength of the $U(N)$ gauge field $A_m$. 
$\tilde{F}^{mn} = \frac{1}{2}\varepsilon^{mnpq} F_{pq}$ is the dual field strength. 
$\Lambda^{\alpha A}, \bar{\Lambda}_{\dot{\alpha} A} 
$ are gauginos, $\varphi_a 
$ are  adjoint scalar fields,
$D_m * = \partial_m * + i g [A_m, *]$ is the gauge covariant derivative,
$g$ is a gauge coupling
constant and $\theta$ is a theta angle.
We denote $T^u$ as the basis of $U(N)$ generators normalized as
${\rm Tr}(T^u T^v)=\kappa\delta^{uv}$ with constant factor $\kappa$.%
\footnote{We use the same normalization for $U(N)$ generators and 
$U(k)$ generators. }
The
(S,A)-deformation term $\mathcal{L}_C$
in (\ref{N4Lag})
is given by \cite{ItNaSa1}
\begin{eqnarray}
\mathcal{L}_C = - \frac{1}{\kappa}
\mathrm{Tr} \left[ igF_{m n} \varphi_a C^{m n a}
- g\varepsilon_{ABCD} \Lambda_{\alpha}^{\ A} \Lambda_{\beta}^{\ B} C^{(\alpha \beta)[CD]}
+ \frac{1}{2} g^2 \varphi_a \varphi_b C_{m n}^{\ \ a} C^{m n b}
\right]+\cdots,
\end{eqnarray}
where the deformation parameters $C^{m n a}$ and $C^{(\alpha \beta)[AB]}$ are defined in (\ref{deformation_parameter}) 
and $\cdots$ 
stands for the $\mathcal{O} (C^3)$ contributions to the Lagrangian $\mathcal{L}$.

The equations of motion are
\begin{eqnarray}
\begin{aligned}
& D^2 \varphi_a - g (\Sigma^a)^{AB} \overline{\Lambda}_{\dot{\alpha} A}
\overline{\Lambda}^{\dot{\alpha}} {}_B - g (\overline{\Sigma}^a)_{AB}
\Lambda^{\alpha A} \Lambda_{\alpha} {}^B + g^2 \Bigl[ \varphi_b,
[\varphi_a, \varphi_b] \Bigr] + \\
& \qquad \qquad + i g F_{m n} C^{m n
a} + g^2 \varphi^b C_{m n} {}^a C^{m n b} = 0, \\
& (\sigma^{m})_{\alpha \dot{\beta}} D_{m} \overline{\Lambda}^{\dot{\beta}}
{}_A - g (\overline{\Sigma}^a)_{AB} [\varphi_a, \Lambda_{\alpha} {}^B]
+ 2 g \varepsilon_{ABCD} \Lambda^{\beta B} C_{(\alpha \beta)} {}^{[CD]} = 0, \\
& (\bar{\sigma}^{m})^{\dot{\alpha} \beta} D_{m} \Lambda_{\beta} {}^A
- g (\Sigma^a)^{AB} [\varphi_a, \overline{\Lambda}^{\dot{\alpha}} {}_B]
= 0, \\
& D_{m} ( F^{m n} + \tilde{F}^{m n} - 2 i g \varphi_a C^{m n
a}) - i g
[\varphi_a, D^{n} \varphi_a] - g (\sigma^{n})_{\alpha \dot{\beta}}
\{\Lambda^{\alpha A}, \overline{\Lambda}^{\dot{\beta}} {}_A\} = 0.
\label{N4eom}
\end{aligned}
\end{eqnarray}
As in the case of the deformed $\mathcal{N} = 2$ super Yang-Mills theory
\cite{ItNaSa2, proItNaSa}, the terms which contain the gauge field
strength and quadratic terms in $C$ in the action
are combined into the perfect square form
$S'$ as
\begin{align}
S'
&=\!\int\!d^{4}x\,\frac{1}{\kappa}\mathrm{Tr}\left[
\frac{1}{2}\bigl(F_{m n}^{(-)}\bigr)^{2}
- igC^{mn a}\varphi_{a}F_{mn}^{(+)}
-\frac{1}{2}g^{2}(C_{mn a}\varphi_{a})^{2}\right]
+ \biggl(
\frac{8\pi^{2}}{g^{2}}+i\theta\biggr)k
\notag\\
&=\!\int\!d^{4}x\,\frac{1}{\kappa}\mathrm{Tr}\left[
\frac{1}{2}\bigl(F_{mn}^{(+)}-igC_{mn a}\varphi_{a}\bigr)^{2}\right]
+\biggl( - \frac{8\pi^{2}}{g^{2}}+i\theta\biggr)k,
\end{align}
where we have defined $F^{(\pm)}_{mn} = \frac{1}{2} (F_{mn} \pm \tilde{F}_{mn})$    
and have used (\ref{C_SD}).
We then obtain the self-dual and anti-self-dual equations for
the gauge field
\begin{eqnarray}
& & F_{m n}^{(-)} = 0, \quad \textrm{for self-dual case,}
\label{n4sd} \\
& & F_{m n}^{(+)} - i g C_{mn a} \varphi_{a} = 0,
\quad \textrm{for anti-self-dual case.} \label{n4asd}
\end{eqnarray}

In the Coulomb branch,
the adjoint scalar fields $\varphi_a$ are able to have VEVs.
We need to expand the solution in
the gauge coupling constant $g$ and solve the equations perturbatively.
However, unlike the $\mathcal{N} = 2$ case, we can not solve the set of equations of motion exactly
even for the $C=0$ case \cite{DoHoKhMa}, so we will expand the field around the
approximate solution of the equations.
The expansion in $g$ is valid when the VEVs of scalar fields are large.
Then the classical action $S$ is expanded in the
gauge coupling constant $g$ as
\begin{equation}
S = \frac{8 \pi^2 |k|}{g^2} + i k \theta + g^0 S^{(0)}_{\mathrm{eff}} +
\mathcal{O} (g^2), \label{classical_g-expansion}
\end{equation}
where $S^{(0)}_{\mathrm{eff}}$
is the instanton effective action
which is expressed by the ADHM moduli.
To calculate the instanton effective action, we need to solve
the equations of motion (\ref{N4eom}) in the instanton background at the leading order in the gauge coupling
constant and write down the solution in terms of the ADHM moduli.
Plugging this solution into the classical action,
we obtain the instanton effective action $S^{(0)}_{\mathrm{eff}}$.

Since the anti-self-dual solution is not deformed as we discuss later,  we investigate the solutions for the self-dual case.
For the self-dual condition (\ref{n4sd}), 
the solution is expanded in the gauge coupling as \cite{DoHoKhMa}
\begin{eqnarray}
\begin{aligned}
A_{m} =& g^{-1} A_{m}^{(0)} + g^1 A_{m}^{(1)} + \cdots ,
\\
\Lambda^A =& g^{-\frac{1}{2}} {\Lambda^{(0)A}} + g^{\frac{3}{2}}
{\Lambda^{(1)A}} + \cdots,
\\
\overline{\Lambda}_A =& g^{\frac{1}{2}} \overline{\Lambda}^{(0)}_A
+ g^{\frac{5}{2}} \overline{\Lambda}^{(1)}_{A} + \cdots,
\\
\varphi_a =& g^0 \varphi_a^{(0)} + g^2 \varphi_a^{(1)} + \cdots.
\label{SD_expansion}
\end{aligned}
\end{eqnarray}
Then the equations of motion in the self-dual background at the leading order are
\begin{eqnarray}
& & (\bar{\sigma}^m)^{\dot{\alpha} \alpha} \nabla_{m} \Lambda_{\alpha}^{(0) A} = 0, \\
& & (\sigma^{m})_{\alpha \dot{\beta}} \nabla_{m}
\overline{\Lambda}^{(0) \dot{\beta}}_A - i (\overline{\Sigma}^a)_{AB}
\bigl[\varphi_a^{(0)}, \Lambda_{\alpha}^{(0) B}\bigr]
+ 2 \varepsilon_{ABCD} \Lambda^{(0) \beta B} C_{(\alpha \beta)}
{}^{[CD]} = 0, \\
& & \nabla^2 \varphi_a^{(0)} - (\overline{\Sigma}^a)_{AB}
\Lambda^{(0) \alpha A} \Lambda^{(0) B}_{\alpha} + i F_{m n}^{(0)}
C^{m n a} = 0,\label{scalar_eq}\\
& & \nabla_{m} (F^{(0) m n} + \tilde{F}^{(0) m n}) = 0,
\label{n4gauge_eq0}
\end{eqnarray}
where $\nabla_m$ denotes the gauge covariant derivative in the self-dual instanton background.
Similar to the
$\mathcal{N} = 2$ case \cite{BiFrFuLe, ItNaSa2, proItNaSa}, we can write the solution in the following form
\begin{eqnarray}
\begin{aligned}
& A^{(0)}_{m} = - i \overline{U} \partial_{m} U, \\
& \Lambda^{(0)A}_{\alpha} = \Lambda_{\alpha} (\mathcal{M}^A)
= \overline{U} (\mathcal{M}^A f \bar{b}_{\alpha} - b_{\alpha} f
\overline{\mathcal{M}}^A) U, \\
& \varphi^{(0)}_a = - \frac{1}{4} (\overline{\Sigma}^a)_{AB}
\overline{U} \mathcal{M}^A f \overline{\mathcal{M}}^B U + \overline{U}
\left(
\begin{array}{cc}
\phi_a^0 & 0 \\
0 & \chi_a \mathbf{1}_{2} + \mathbf{1}_{k} C_a
\end{array}
\right)
U,
\label{ADHM_sol}
\end{aligned}
\end{eqnarray}
where $\phi_a^0$ are the VEVs of the adjoint scalar fields $\varphi_a$.
$C_a$ is the 2$\times$2 matrix whose components are
$(C_a)_{\alpha}{}^{\beta} = (\sigma^{m n})_{\alpha} {}^{\beta} C_{m n a}$.
$\chi_a$ should satisfy the equation
\begin{equation}
\mathbf{L} \chi_a =
\frac{1}{4} (\overline{\Sigma}^a)_{AB}
\bar{\mathcal{M}}^A \mathcal{M}^B
+ \bar{w}^{\dot{\alpha}}
\phi_a^0 w_{\dot{\alpha}}
+ C^{m n a} [a'_{m}, a'_{n}],
\end{equation}
where $\mathbf{L}$ is defined in (\ref{definition_bfL}).
We present the derivation of these solutions in Appendix B.
As in the $\mathcal{N} = 2$ case, the self-dual condition $F_{m n}^{(0)(-)} = 0$ 
is consistent with the equation of motion (\ref{n4gauge_eq0}).
Note that we do not need to find the solution for the anti-chiral fermion $\bar{\Lambda}^{(0)}$ since it contributes
to the classical action as the subleading order in the gauge coupling constant.

Substituting the expansion (\ref{SD_expansion}) back into the
classical action, the instanton effective action $S_{\mathrm{eff}}^{(0)}$ in (\ref{classical_g-expansion})
is
\begin{eqnarray}
S^{(0)}_{\mathrm{eff}}
&=& \int \! d^4 x \frac{1}{\kappa} \partial_m
\mathrm{tr}_k
\left[
\frac{1}{2} \varphi^{(0)}_a \nabla_m \varphi^{(0)}_a
\right]  \nonumber
\\
& & + \int \! d^4 x \ \frac{1}{\kappa}
\mathrm{tr}_k
\left[
- \frac{1}{2} (\overline{\Sigma}^a)_{AB}
\Lambda^{(0) \alpha A} [\varphi_a^{(0)}, \Lambda_{\alpha}^{(0) B}]
- \frac{1}{2} (\overline{\Sigma}^a)_{AB}
C^{(\alpha \beta) a} \Lambda^{(0)A}_{\alpha} \Lambda_{\beta}^{(0)B}
\right.
\nonumber
\\
& & \qquad \qquad \qquad \left.
+ \frac{1}{4} (\overline{\Sigma}^a)_{AB}
\Lambda^{(0) \alpha A} [\varphi_a^{(0)}, \Lambda_{\alpha}^{(0) B}] 
- \frac{i}{2} \varphi_a^{(0)} F^{(0)}_{mn} C^{mna}
\right],
\label{N4inst}
\end{eqnarray}
where we have decomposed the Yukawa term for later convenience.
The first term in (\ref{N4inst}) is easily computed as in the case of the
undeformed theory. After plugging the solution of the scalar field in (\ref{ADHM_sol}) into the
first term in the above expression, we find that the $C$-dependent parts are canceled out
and the result is
\begin{eqnarray}
\frac{2\pi^2}{\kappa} \mathrm{tr}_k
\left[
\frac{1}{4}
(\overline{\Sigma}^a)_{AB}
\bar{\mu}^A \phi_a^0 \mu^B - \bar{w}_{\dot{\alpha}} \phi_a^0 \phi_a^0 w^{\dot{\alpha}}
+ \bar{w}_{\dot{\alpha}} \phi_a^0 w^{\dot{\alpha}} \chi_a
\right].
\end{eqnarray}
Next, let us consider the second term in (\ref{N4inst}).
To compute this term, we use the following relations
\begin{eqnarray}
- \frac{1}{2} \Lambda^{(0)\alpha} (\mathcal{M}^A)
[\varphi_a^{(0)}, \Lambda^{(0)}_{\alpha} (\mathcal{M}^B)]
&=& - \frac{1}{2}
(\sigma^m)_{\alpha \dot{\alpha}} \Lambda^{(0) \alpha} (\mathcal{M}^A)
\nabla_m \bar{\psi}^{\dot{\alpha}} {}_A
- \frac{1}{2} \Lambda^{(0) \alpha} (\mathcal{M}^A) \Lambda^{(0)}_{\alpha}
(\mathcal{N}_A)
\nonumber \\
& &
- \frac{1}{2} \Lambda^{(0)}_{\alpha} (\mathcal{M}^A) \Xi^{\alpha} {}_A,
\label{Yukawa_decomposition}
\end{eqnarray}
where
\begin{eqnarray}
\begin{aligned}
& \bar{\psi}_A^{\dot{\alpha}} =
\bar{\psi}_A^{(1) \dot{\alpha}}
+ \bar{\psi}_A^{(2)\dot{\alpha}}
+ \bar{\psi}_A^{(3)\dot{\alpha}}, \\
& \bar{\psi}_{\dot{\alpha} A}^{(1)}
=  \frac{1}{4} \varepsilon_{ABCD} \overline{U}
\mathcal{M}^B f \overline{\Delta}_{\dot{\alpha}} \mathcal{M}^C
f \overline{\mathcal{M}}^D U, \\
& \bar{\psi}_{\dot{\alpha} A}^{(2)} =
\frac{1}{2} (\overline{\Sigma}^a)_{AB} \overline{U}
\left\{
- \mathcal{M}^B f \overline{\Delta}_{\dot{\alpha}} M
+ M \Delta_{\dot{\alpha}} f \overline{\mathcal{M}}^B
\right\} U, \\
& \bar{\psi}_{\dot{\alpha} A}^{(3)} = \overline{U} Q_{\dot{\alpha}A} U, \\
& \Xi_{\alpha A} = (\overline{\Sigma}^a)_{AB} (C^a)_{\alpha} {}^{\beta}
\Lambda_{\beta} (\mathcal{M}^B),
\end{aligned}
\end{eqnarray}
and the matrices $M$, $Q_{\dot{\alpha} A}$ and $\mathcal{N}$ are given by
\begin{eqnarray}
M_{\lambda} {}^{\mu} &=& M_{(u+ i \alpha)} {}^{(v + j \beta)}
=
\left(
\begin{array}{cc}
(\varphi_a^0)_u {}^v & 0 \\
0 & (\chi_a)_i {}^j \delta_{\alpha} {}^{\beta}
+ \delta_i {}^j (C_a)_{\alpha} {}^{\beta}
\end{array}
\right), \\
Q_{\dot{\alpha}A} &=&
\left(
\begin{array}{cc}
0 & 0 \\
0 & (\mathcal{G}_{\dot{\alpha} A})_{ij} \delta_{\alpha} {}^{\beta}
\end{array}
\right), 
\\
\mathcal{N}_A &=&
(\overline{\Sigma}^a)_{AB} \left[
M \mathcal{M}^B - \mathcal{M}^B \chi_a
\right]
+ 2
\left(
\begin{array}{cc}
0 & 0 \\
0 & \mathcal{G}^{\dot{\alpha}} {}_A
\end{array}
\right)
a_{\dot{\alpha}}
- 2 a_{\dot{\alpha}} \mathcal{G}^{\dot{\alpha}} {}_A, \\
\overline{\mathcal{N}}_A &=&
(\overline{\Sigma}^a)_{AB} \left[
- \overline{\mathcal{M}}^B M + \chi_a \overline{\mathcal{M}}^B
\right]
+ 2 \bar{a}_{\dot{\alpha}}
\left(
\begin{array}{cc}
0 & 0 \\
0 & \overline{\mathcal{G}}^{\dot{\alpha}} {}_A
\end{array}
\right)
- 2 \overline{\mathcal{G}}^{\dot{\alpha}} {}_A \bar{a}_{\dot{\alpha}}.
\end{eqnarray}
Here $\mathcal{G}_{\dot{\alpha}A}$ is a constant anti-Hermitian matrix and is determined so that $\mathcal{N}_{A}$ satisfies
the fermionic ADHM constraint (\ref{fADHM}).
The last term in (\ref{Yukawa_decomposition}) cancels the third term
in (\ref{N4inst}) while the first term is rewritten as the total derivative term and does not contribute to the
instanton effective action.
The second term in (\ref{Yukawa_decomposition}) is evaluated by Corrigan's
inner product formula \cite{DoHoKhMaVa},
which is expressed as
\begin{eqnarray}
& & \int \! d^4 x \ \frac{1}{\kappa}
\mathrm{tr}_k
\left[
- \frac{1}{2} \Lambda^{ (0) \alpha} (\mathcal{M}^A)
\Lambda^{(0)}_{\alpha} (\mathcal{N}_A)
\right]
\nonumber \\
&=& \frac{2\pi^2}{\kappa}
\mathrm{tr}_k
\left[
\frac{1}{2} (\overline{\Sigma}^a)_{AB}
\left(
\bar{\mu}^A \phi^0_a \mu^B - \bar{\mu}^A \mu^B \chi_a
- \mathcal{M}^{\prime \alpha A} \mathcal{M}^{\prime}_{\alpha} {}^B \chi_a
\right)
- \frac{1}{4} (\overline{\Sigma}^a)_{AB} C^{(\alpha \beta) a}
\mathcal{M}^{\prime}_{\alpha} {}^A
\mathcal{M}^{\prime}_{\beta} {}^B
\right].
\nonumber \\
\end{eqnarray}
Let us calculate the fourth and the last terms in the equation (\ref{N4inst}).
To calculate these terms, we decompose the scalar field
\begin{eqnarray}
\varphi_a^{(0)} = \varphi^{(0)}_{\mathcal{M}, a}
+ \varphi^{(0)}_{\phi,a} +
\varphi^{(0)}_{C,a}\,,
\end{eqnarray}
where
\begin{eqnarray}
\varphi^{(0)}_{\mathcal{M}, a}
&=&
- \frac{1}{4} (\overline{\Sigma}^a)_{AB}
\overline{U} \mathcal{M}^A f \overline{\mathcal{M}}^B U
+ \overline{U}
\left(
\begin{array}{cc}
0 & 0 \\
0 & \chi_{\mathcal{M},a} \mathbf{1}_2
\end{array}
\right)
U, \\
\varphi^{(0)}_{\phi,a}
&=&
\overline{U}
\left(
\begin{array}{cc}
\phi^0_a & 0 \\
0 & \chi_{\phi,a} \mathbf{1}_2
\end{array}
\right)
U, \\
\varphi^{(0)}_{C,a}
&=&
\overline{U}
\left(
\begin{array}{cc}
0 & 0 \\
0 & \chi_{C,a} \mathbf{1}_2
+ \mathbf{1}_k C_a
\end{array}
\right)
U.
\end{eqnarray}
We also decompose $\chi_a$ as
\begin{equation}
\chi_a = \chi_{\mathcal{M},a}
+ \chi_{\phi,a} + \chi_{C,a}\,,
\end{equation}
where
\begin{eqnarray}
\chi_{\mathcal{M},a} &=&
\mathbf{L}^{-1}
\left(
\frac{1}{4} (\overline{\Sigma}^a)_{AB}
\bigl(\bar{\mu}^A \mu^B + \mathcal{M}^{\prime \alpha A}
\mathcal{M}^{\prime}_{\alpha} {}^B \bigr)
\right), \\
\chi_{\phi,a} &=& \mathbf{L}^{-1}
(\bar{w}^{\dot{\alpha}} \phi^0_a w_{\dot{\alpha}}), \\
\chi_{C,a} &=& \mathbf{L}^{-1}
\left(
C^{mna} [a'_m, a'_n]
\right).
\end{eqnarray}
Then, we can rewrite the sum of the fourth and the last terms in (\ref{N4inst}) as
\begin{eqnarray}
& & \int \! d^4 x \ \frac{1}{\kappa}
\mathrm{tr}_k
\left[
\frac{1}{4} (\overline{\Sigma}^a)_{AB}
\Lambda^{(0) \alpha A}
[\varphi^{(0)}_{\mathcal{M},a} + \varphi^{(0)}_{\phi,a},
\Lambda^{(0)B}_{\alpha}]
- \frac{1}{2} (\overline{\Sigma}^a)_{AB}
\varphi^{(0)}_{C,a} \Lambda^{(0) \alpha A}
\Lambda^{(0)}_{\alpha} {}^B 
\right.
\nonumber
\\
& & \qquad \qquad \left.
- \frac{i}{2} C^{mna} \varphi^{(0)}_{\mathcal{M},a} F^{(0)}_{mn}
- \frac{i}{2} C^{mna}
(\varphi^{(0)}_{\phi,a} + \varphi^{(0)}_{C,a})
F^{(0)}_{mn}
\right].
\label{N4inst2}
\end{eqnarray}
The first term in the above integral
is independent of $C$ and easily evaluated as
\begin{eqnarray}
\frac{2\pi^2}{\kappa}
\mathrm{tr}_k
\biggl[
- \frac{1}{4}
(\overline{\Sigma}^a)_{AB}
\Bigr(
\bar{\mu}^A \phi^0_a \mu^B
- \bigl(\bar{\mu}^A \mu^B
+ \mathcal{M}^{\prime \alpha A}
\mathcal{M}^{\prime}_{\alpha} {}^B \bigr)
(\chi_{\mathcal{M},a} + \chi_{\phi,a})
\Bigr)
\biggr].
\end{eqnarray}
Since $\varphi^{(0)}_{\mathcal{M},a}$ and $\varphi^{(0)}_{C,a}$ satisfy
\begin{eqnarray}
\nabla^2 \varphi^{(0)}_{\mathcal{M},a}
&=& (\overline{\Sigma}^a)_{AB} \Lambda^{(0)\alpha A}
\Lambda_{\alpha}^{(0) B}, \\
\nabla^2 \varphi^{(0)}_{C,a}
&=& - i C^{mna} F_{mn}^{(0)},
\end{eqnarray}
the second and third terms in the integral  (\ref{N4inst2}) become the total derivatives and are evaluated as
\begin{eqnarray}
- \frac{1}{2} \lim_{|x| \to \infty}
2 \pi^2 |x|^3
\frac{x^m}{|x|}
\mathrm{tr}_k
\left[
\varphi^{(0)}_{C,a} \nabla_m \varphi^{(0)}_{\mathcal{M},a}
- \varphi^{(0)}_{\mathcal{M},a} \nabla_m \varphi^{(0)}_{C,a}
\right].
\end{eqnarray}
We find that this term vanishes at the boundary $|x| \to \infty$.
Finally, we focus on the
last term in (\ref{N4inst2}).
Since we have the relation
\begin{eqnarray}
\varphi^{(0)}_{\phi,a} + \varphi^{(0)}_{C,a}
&=&
\overline{U}
\left(
\begin{array}{cc}
\phi_a^0 & 0 \\
0 & 
(\chi_{\phi,a} + \chi_{C,a})\mathbf{1}_2 + \mathbf{1}_k C^a
\end{array}
\right)
U
\equiv \overline{U} \hat{M} U, %
\end{eqnarray}
the last term in (\ref{N4inst2}) is rewritten as
\begin{eqnarray}
-4 (C^a)^{\alpha} {}_{\beta}
\int \! d^4 x \ \frac{1}{\kappa}
\mathrm{tr}_k
\left[
\overline{U} \hat{M} \mathcal{P} b_{\alpha} f \bar{b}^{\beta} U
\right],
\label{UMPbfbU}
\end{eqnarray}
where $\mathcal{P}_{\lambda} {}^{\mu}$ is the projection operator defined by
$\mathcal{P}_{\lambda} {}^{\mu} \equiv U_{\lambda u} \bar{U}_u {}^{\mu}
= \delta^{\mu} {}_{\lambda} - \Delta_{\lambda i \dot{\alpha}} f_{ij} \bar{\Delta}_j {}^{\dot{\alpha} \mu}$.
The term (\ref{UMPbfbU}) can be evaluated as in the same way in $\mathcal{N} = 2$ case \cite{ItNaSa2}.
The result is
\begin{eqnarray}
\frac{2\pi^2}{\kappa}
\mathrm{tr}_k
\left[
\frac{1}{4}
C^{mna} C_{mna} \bar{w}^{\dot{\alpha}} w_{\dot{\alpha}}
- C^{mna} ( \chi_{\phi,a} + \chi_{C,a}) [a'_m, a'_n]
\right].
\end{eqnarray}
From these results $S_{\mathrm{eff}}^{(0)}$ can be written in terms of ADHM moduli as follows
\begin{eqnarray}
S^{(0)}_{\mathrm{eff}}
&=& \frac{2\pi^2}{\kappa}
\mathrm{tr}_k
\left[
\frac{1}{2} (\overline{\Sigma}^a)_{AB}
\bar{\mu}^A \phi^0_a \mu^B
- \bar{w}_{\dot{\alpha}} \phi^0_a \phi^0_a w^{\dot{\alpha}}
- \chi_a \mathbf{L} \chi_a
\right. \nonumber \\
& & \qquad \qquad
\left.
- \frac{1}{4} (\overline{\Sigma}^a)_{AB} C^{(\alpha \beta) a}
\mathcal{M}^{\prime}_{\alpha} {}^A \mathcal{M}^{\prime}_{\beta} {}^B
+ \frac{1}{4} C^{mna} C_{mna} \bar{w}^{\dot{\alpha}} w_{\dot{\alpha}}
\right].
\end{eqnarray}
In the  above instanton effective action, the ${\cal N}=4$ ADHM moduli obey the ADHM constraints (\ref{ADHM2}) and
(\ref{fADHM}).
We introduce auxiliary fields $D^{c}$ 
and $\bar{\psi}^{\dot{\alpha}}_{A}$  for these constraints and add the Lagrange multiplier terms to the
effective action.
Then we can show that the $S^{(0)}_{\mathrm{eff}}$ can be obtained from the following action by integrating out the auxiliary fields:
\begin{eqnarray}
\tilde{S}^{(0)}_{\mathrm{eff}}
&=& \frac{2\pi^2}{\kappa}
\mathrm{tr}_k
\left[
- [\chi_a, a'_m]^2
+ (w_{\dot{\alpha}} \chi_a - \phi^0_a w_{\dot{\alpha}})
(\chi_a \bar{w}^{\dot{\alpha}} - \bar{w}^{\dot{\alpha}} \phi^0_a)
\right.
\nonumber \\
& & \qquad
+ \frac{1}{2} (\overline{\Sigma}^a)_{AB}
\bar{\mu}^A (- \chi_a \mu^B + \phi^0_a \mu^B)
- \frac{1}{2} (\overline{\Sigma}^a)_{AB}
\mathcal{M}^{\prime \alpha A} \mathcal{M}^{\prime}_{\alpha} {}^B \chi_a
\nonumber \\
& & \qquad
\left.
- \frac{1}{4} (\overline{\Sigma}^a)_{AB} C^{(\alpha \beta) a}
\mathcal{M}^{\prime}_{\alpha} {}^A \mathcal{M}^{\prime}_{\beta} {}^B
- 2 C^{mna} [a'_m, a'_n] \chi_a
+ \frac{1}{4} C^{mna} C_{mna} \bar{w}^{\dot{\alpha}} w_{\dot{\alpha}}
\right] \nonumber \\
& & + S_{\textrm{ADHM}}.
\label{Inst_eff_D3}
\end{eqnarray}
We also call $\tilde{S}^{(0)}_{\mathrm{eff}}$ the instanton effective action.
The result does not agree with the string theory calculation (\ref{deformed_inst_eff_action}) at the
second order in the deformation parameter. This is the conceivable
result since the (S,A)-deformed gauge theory does not break the
translational invariance and the mass term for the position moduli $a'_m$ is not allowed.
To resolve this discrepancy, let us introduce the following term
and improve the (S,A)-deformed theory:
\begin{eqnarray}
\delta S
= \frac{g^2}{2\kappa} \int \! d^4 x \mathrm{Tr}
\left[
C_{mpa} C_{nqa} x^p x^q F^{mr} F^{n} {}_r
\right].
\label{FFCC}
\end{eqnarray}
This term does not provide any modifications to the equations of
motion at the leading order in $g$ and hence we can use the same solution (\ref{ADHM_sol}) to calculate
the instanton effective action. The additional contribution to the
instanton effective action from the term (\ref{FFCC}) is easily evaluated as
\begin{eqnarray}
\frac{g^2}{2\kappa} \int \! d^4 x \mathrm{Tr}
\left[
C_{mpa} C_{nqa} x^p x^q F^{(0)mr} F^{(0)n} {}_r
\right]
=
\frac{2\pi^2}{\kappa} \mathrm{tr}_k
\biggl[
- \frac{1}{4} (a'_m a^{\prime m} + \bar{w}^{\dot{\alpha}} w_{\dot{\alpha}})
\biggr]
C^{pqa} C_{pqa}.
\end{eqnarray}
After adding this contribution to (\ref{Inst_eff_D3}),
the improved instanton effective action becomes the same as (\ref{deformed_inst_eff_action}).

Before going to the next section, let us comment on the anti-self-dual case. 
For the anti-self-dual case,
it is easy to see that these 
equations are not deformed by $C$. The instanton effective action still
contains a term $F_{mn}^{(0)} \varphi_a^{(0)} C^{m n a}$, but
this term vanishes due to the self-dual condition of the background (\ref{C_SD}).
Therefore the instanton effective action does not receive any
deformation effect for the anti-self-dual case.

\section{$\Omega$-background deformation of $\mathcal{N} = 4$ super Yang-Mills theory}
In this section,
we discuss the relation between the deformed instanton effective action obtained in section 2, 3 and the one
derived from $\mathcal{N} = 4$ super Yang-Mills theory in the $\Omega$-background.
We will find that the deformed instanton effective action (\ref{deformed_inst_eff_action}) is interpreted as 
the one calculated in $\mathcal{N} = 4 $ super Yang-Mills theory in the $\Omega$-background.
Similar to the $\Omega$-background deformation of four-dimensional
$\mathcal{N} = 2$ theories which is given by the dimensional
reduction of the six-dimensional theory \cite{Ne}, the $\Omega$-background deformation of
$\mathcal{N} = 4$ theory can be obtained by the dimensional reduction of the
ten-dimensional $\mathcal{N} = 1$ super Yang-Mills theory in the
non-trivial metric
\begin{eqnarray}
d s^2_{10} = (d x^a)^2 + (d x^{m} + \Omega_{m a} d x^a)^2,
\qquad
\Omega_{ma} = \Omega_{mna} x^n.
\label{omega_metric}
\end{eqnarray}
Here $\Omega_{mna}=-\Omega_{nma}$.
The four and six-dimensional indices $m$ and $a$ are raised and lowered by flat metric.
The Lagrangian of $\mathcal{N}=1$ super Yang-Mills theory in the metric (\ref{omega_metric}) is given by 
\begin{eqnarray}
\mathcal{L} (\Omega) = \frac{1}{\kappa g^2} \mathrm{Tr} \sqrt{- g} \left[ - \frac{1}{4}
F_{\mathcal{M N}} F_{\mathcal{PQ}} g^{\mathcal{MP}} g^{\mathcal{NQ}} - \frac{i}{2} \overline{\Psi} e^{\mathcal{M}} {}_M \Gamma^M
\mathcal{D}_{\mathcal{M}} \Psi
\right],
\end{eqnarray}
where $F_{\mathcal{M N}} = \partial_{\mathcal{M}} A_{\mathcal{N}} - \partial_\mathcal{N} A_{\mathcal{M}} + i [A_{\mathcal{M}}, A_{\mathcal{N}}]$
is the field strength of the gauge field $A_{\mathcal{M}}$ and
$\Psi$ is the ten-dimensional Majorana-Weyl spinor \cite{BrScSc}.
$\Gamma^M$ is the ten-dimensional gamma matrix satisfying $\{\Gamma^M,
\Gamma^N \} = - 2 \eta^{MN}$. $\mathcal{M, N, P, Q} = 0, \ldots, 9$ are curved indices
while $M, N, \cdots = 0, \ldots, 9$ are local Lorentz indices. $e^{\mathcal{M}} {}_M$ is the
ten-dimensional vielbein and the covariant derivative is defined by
\begin{eqnarray}
\mathcal{D}_{\mathcal{M}} = D_{\mathcal{M}} - \frac{1}{2} \omega_{\mathcal{M} MN} \Gamma^{MN},
\end{eqnarray}
where $D_{\mathcal{M}} * = \partial_{\mathcal{M}} * + i [A_{\mathcal{M}}, *]
$ and $\omega_{\mathcal{M} MN}$ is the spin connection. The Lorentz generator
is defined as $\Gamma^{MN} = \frac{1}{4}[\Gamma^M, \Gamma^N]$.
We require that the $\Omega$-background spacetime is flat. As we will see later, this requirement ensures the existence of the supersymmetry for the instanton effective action.
We find that the flatness condition of the spacetime needs 
\begin{eqnarray}
\Omega^{m n} {}_a \Omega_{n p b}
- \Omega^{m n} {}_b \Omega_{n p a} 
= 0.
\label{flatness_condition}
\end{eqnarray}
This is the natural generalization of the flatness condition in the
six-dimensional $\Omega$-background \cite{Ne}.
If the flatness condition (\ref{flatness_condition}) is satisfied,
the only non-zero component in the spin connection for the metric (\ref{omega_metric})
is given as
\begin{eqnarray}
\omega_{\mathcal{A} mn} = - \Omega_{mn \mathcal{A}}.
\end{eqnarray}
After the dimensional reduction to four dimensions,
we obtain the deformed $\mathcal{N} = 4$ super Yang-Mills theory.
The action, which is expanded
up to the second order in $\Omega_{mna}$, has the form
$\mathcal{L} (\Omega) = \mathcal{L}_0 + \delta \mathcal{L} (\Omega)$,
where $\mathcal{L}_0$ is the 
Lagrangian of  $\mathcal{N} = 4$ super Yang-Mills theory 
(\ref{N4SYM}) and $\delta \mathcal{L} (\Omega)$ is the 
term which depends on 
the deformation parameter $\Omega_{mna}$.
This is given by
\begin{eqnarray}
\delta \mathcal{L} (\Omega)
&=& \frac{1}{\kappa} \mathrm{Tr} \left[ g F_{m n} D^{m} \varphi_a \Omega^{n} {}_a
+ i g^2 D_{m} \varphi_a
[\varphi_b, \varphi_a] \Omega^{m} {}_b
+ \frac{g^2}{2} F_{m p}
F^{p} {}_{n} \Omega^{m} {}_a \Omega^{n} {}_a \right.
\nonumber \\
& &
+ \frac{g^2}{2} D_{m} \varphi_b D_{n} \varphi_a
\Omega^{m} {}_a \Omega^{n} {}_b
- \frac{ig^3}{2} F_{m n} [\varphi_a, \varphi_b] \Omega^{m} {}_a
\Omega^{n} {}_b
- \frac{g^2}{2} D_{m} \varphi_a D_{n} \varphi_a \Omega^{m} {}_b
\Omega^{n} {}_b \nonumber \\
& & + \frac{i g}{2} \Omega^{m} {}_a
\left[
(\overline{\Sigma}^a)_{AB} \Lambda^{\alpha A} D_{m} \Lambda_{\alpha} {}^B
+ (\Sigma^a)^{AB} \overline{\Lambda}_{\dot{\alpha} A} D_{m}
\overline{\Lambda}^{\dot{\alpha}} {}_B
\right] \nonumber \\
& & \left.
- \frac{i g}{4} \Omega_{m n a} \left[
(\overline{\Sigma}^a)_{AB} \Lambda^{\alpha A} (\sigma^{m n})_{\alpha}
{}^{\beta} \Lambda_{\beta} {}^B
+ (\Sigma^a)^{AB} \overline{\Lambda}_{\dot{\alpha} A}
(\bar{\sigma}^{m n})^{\dot{\alpha}} {}_{\dot{\beta}}
\overline{\Lambda}^{\dot{\beta}} {}_B \right] \right] +
\mathcal{O} (\Omega^3). \nonumber \\
\end{eqnarray}
Here we have rescaled all the fields and the deformation parameter as
$(A_{m}, \varphi_a, \Lambda_{\alpha} {}^A, \Omega_{ma}) \to g (A_{m}, \varphi_a, \Lambda_{\alpha} {}^A, \Omega_{ma}) $
so that one can see clearly the power
of the gauge coupling constant in each term.

Now we are interested in the instanton effective action of this deformed theory.
Assuming the self-dual condition of the gauge field and using the gauge
coupling expansion of the solution (\ref{SD_expansion}),
the equations of motion at the leading order in $g$ are given by
\begin{eqnarray}
\begin{aligned}
& i (\bar{\sigma}^{m})^{\dot{\alpha} \alpha } \nabla_{m} \Lambda^{(0)}_{\alpha}
{}^A = 0, 
\\
& i (\sigma^{m})_{\alpha \dot{\alpha}} \nabla_{m}
\overline{\Lambda}^{ (0) \dot{\alpha}} {}_A
+ (\Sigma^a)_{AB} [\varphi^{(0)}_a, \Lambda^{(0)}_{\alpha} {}^B] \\
& \qquad \qquad - i \Omega^{m} {}_a (\overline{\Sigma}^a)_{AB}
\nabla_{m} \Lambda^{(0)}_{\alpha} {}^B
+ \frac{i}{2} \Omega_{m n} {}^a (\overline{\Sigma}^a)_{AB}
(\sigma^{m n})_{\alpha} {}^{\beta} \Lambda^{(0)}_{\beta} {}^B = 0, 
\\
& \nabla^2 \varphi_a^{(0)} + F_{m n}^{(0)} \Omega^{m n a}
- (\nabla^{m} F_{m n}^{(0)}) \Omega^{n a} =
(\overline{\Sigma}^a)_{AB} \Lambda_{\alpha}^{(0)A} \Lambda^{(0) \alpha B},
\\
& \nabla^{m} (F^{(0)}_{m n} + \tilde{F}^{(0)}_{m n})
= 0. \label{omega_F_eom}
\end{aligned}
\end{eqnarray}
If we identify the $\Omega$-background and (S,A)-background parameters through the relation
\begin{eqnarray}
\Omega^{mna} = i C^{mna},
\label{Omega_C}
\end{eqnarray}
the equations of motion for $A_{m}^{(0)}, \varphi_a^{(0)}, \Lambda^{(0)}$ are precisely
equivalent to the one in the (S,A)-deformed $\mathcal{N} = 4$ super
Yang-Mills theory (\ref{N4eom})
because of the self-dual condition of the gauge field and the deformation parameters (\ref{C_SD}).
On the other hand, the
equation for $\overline{\Lambda}^{(0)}$ is different from the (S,A)-deformed
theory. However, this does not matter when we discuss the instanton
effective action because the contributions of $\bar{\Lambda}^{(0)}$
to the instanton effective action is subleading order in $g$ and have no
effects in the semi-classical approximation. Explicitly, after expanding the spacetime action around the instanton background, 
the terms at the order $g^0$ are given by
\begin{eqnarray}
S^{(0)}_{\mathrm{eff}} (\Omega) \!\!\! &=& \!\!\!
\int \! d^4 x \ \frac{1}{\kappa}
\mathrm{Tr} \left[ \frac{1}{2} \nabla_{m} \varphi^{(0)}_a
\nabla^{m} \varphi^{(0)}_a - \frac{1}{2} (\bar{\Sigma}^a)_{AB}
\Lambda^{(0) \alpha A} [\varphi_a^{(0)}, \Lambda_{\alpha}^{(0) B}]
- F_{m n}^{(0)} \varphi^{(0)}_a \Omega^{m n a}
\right. \nonumber \\
& & \qquad \left.
- \frac{1}{2} F^{(0) m p} F^{(0)}_{p} {}^{n} \Omega_{m} {}^a
\Omega_{n} {}^a
+ \frac{i}{4} \Omega_{m n a} (\overline{\Sigma}^a)_{AB}
\Lambda^{(0) \alpha A} (\sigma^{m n})_{\alpha \beta} \Lambda^{ (0) \beta B} \right].
\label{Omega_inst_action}
\end{eqnarray}
There is no $\bar{\Lambda}^{(0)}$-dependence in (\ref{Omega_inst_action}). 
Here we have used
\begin{eqnarray}
\Omega^{m} {}_a (\overline{\Sigma}^a)_{AB}
\Lambda^{(0) \alpha A} \nabla_{m} \Lambda^{(0)}_{\alpha} {}^B
&=& - \Omega^{m n} {}_a (\overline{\Sigma}^a)_{AB}
\Lambda^{(0) \alpha A} (\sigma_{m n})_{\alpha \beta}
\Lambda^{(0) \beta B} \nonumber \\
& & \qquad \qquad + \textrm{(total derivative)},
\end{eqnarray}
and the total derivative part does not contribute to
$S^{(0)}_{\mathrm{eff}}(\Omega)$ in the instanton background.
Therefore,
the instanton effective action for the $\mathcal{N} = 4$
super Yang-Mills theory in the $\Omega$-background (\ref{Omega_inst_action}) is equivalent to the
one for the $\mathcal{N} = 4$ (S,A)-deformed Yang-Mills theory (\ref{N4inst})
with the improvement term (\ref{FFCC}) under the identification (\ref{Omega_C}).

\section{(A,S)-deformed instanton effective action}
In this section, we introduce the (A,S)-type background and study  the deformed instanton effective action.
In \cite{ItNaSa1}, we imposed the self-dual condition for the internal indices of the (A,S)-type background.
In this paper we will consider the anti-self-dual condition, which induces the holomorphic deformation of
the effective action.
The vertex operator corresponding to this background
in the $(-1/2,-1/2)$ picture is given as
\begin{eqnarray}
V^{(-1/2,-1/2)}_{\mathcal{F}} (z,\bar{z}) =
(2 \pi \alpha') \mathcal{F}^{[\dot{\alpha} \dot{\beta}]} {}_{(AB)} S_{\dot{\alpha}} S^{A}
e^{- \frac{1}{2} \phi} (z) S_{\dot{\beta}} S^B e^{- \frac{1}{2} \phi} (\bar{z}).
\end{eqnarray}
As in the case of the (S,A)-background,
the non-zero amplitudes which include one (A,S)-background vertex operator
are found to be
\begin{eqnarray}
\langle \! \langle
V^{(-1/2)}_{\mathcal{M}'}
V^{(-1/2)}_{\mathcal{M}'}
V^{(-1/2,-1/2)}_{\mathcal{F}}
\rangle \! \rangle,
\qquad
\langle \! \langle
V^{(-1/2)}_{\bar{\mu}}
V^{(-1/2)}_{\mu}
V^{(-1/2,-1/2)}_{\mathcal{F}}
\rangle \! \rangle.
\end{eqnarray}
These amplitudes are evaluated in appendix A.
After taking the zero-slope limit, these amplitudes are reproduced by the following interactions on the D$(-1)$-branes
\begin{eqnarray}
\delta S_{\mathrm{(A,S)}} &=& - \frac{2\pi^2}{\kappa}
\mathrm{tr}_k
\left[
\left(
2 \bar{\mu}^A \mu^B + \mathcal{M}^{\prime \alpha A} \mathcal{M}^{\prime}
{}_{\alpha} {}^B
\right) m_{(AB)}
\right],
\label{AS_string}
\end{eqnarray}
where we have defined the deformation parameter
\begin{eqnarray}
m_{(AB)} \equiv \pi i (2\pi \alpha')^{\frac{1}{2}}
\mathcal{F}^{[\dot{\alpha} \dot{\beta}]} {}_{(AB)}
\varepsilon_{\dot{\alpha} \dot{\beta}}.
\end{eqnarray}
This result is also derived from the field theory side.
The (A,S)-background induces new interaction terms on the D3-branes
giving the deformed $\mathcal{N} = 4$ super Yang-Mills theory.
After the Wick rotation the action of the (A,S)-deformed $\mathcal{N} = 4$ super Yang-Mills theory is  \cite{ItNaSa1}
\begin{eqnarray}
\hat{S}=  \int \! d^4 x \ ( \mathcal{L}_0 + \delta \mathcal{L}_{\mathrm{(A,S)}}),
\end{eqnarray}
where $\mathcal{L}_0$ is the Lagrangian of 
$\mathcal{N} = 4$ super Yang-Mills 
theory 
(\ref{N4SYM})
and $\delta \mathcal{L}_{\mathrm{(A,S)}}$ is the induced interaction term given by
\begin{eqnarray}
\delta \mathcal{L}_{\mathrm{(A,S)}}
&=& \frac{1}{\kappa} \mathrm{Tr}
\biggl[
- g^2 (\Sigma^a \overline{\Sigma}^b \Sigma^c)^{AB}
\varphi_a \varphi_b \varphi_c m_{(AB)}
+ 2 g \Lambda^{\alpha A} \Lambda_{\alpha} {}^B m_{(AB)}
\nonumber \\
& &
- \frac{1}{4} g^2
(\Sigma^a \overline{\Sigma}^b \Sigma^c)^{AB}
(\Sigma^a \overline{\Sigma}^b \Sigma^d)^{CD}
\varphi_c \varphi_d m_{(AB)} m_{(CD)}
\biggr].
\label{AS_deformation_term}
\end{eqnarray}
Since there are no gauge field interactions in (\ref{AS_deformation_term}),
the (anti-)self-dual condition for the gauge fields $F^{(\pm)}_{mn} = 0$ is not modified by the
(A,S)-background. As discussed in section 2, we expand the fields by the gauge coupling
constant $g$ and solve the equations of motion at the leading order in $g$.
When we consider the anti-self-dual condition, we find that all the terms in (\ref{AS_deformation_term}) are
subleading order in $g$. Therefore there are no (A,S)-background corrections to the instanton effective action for the anti-self-dual case.
On the other hand, when we consider the self-dual case and expansion (\ref{SD_expansion}), we find that the bi-fermion term in (\ref{AS_deformation_term})
contributes to the instanton effective action while
other terms are subleading order in $g$.
This bi-fermion term is recognized as the mass term for the chiral fermion $\Lambda$.
It is known that the chiral mass term 
is enough to see the effects of the instanton corrections to holomorphic quantities \cite{DoHoKhMa}.

The equations of motion 
at the leading order 
for all the fields except $\bar{\Lambda}^{(0)}$ are the
same with that of the undeformed $\mathcal{N}=4$ super Yang-Mills theory.
We can use the solution for the ordinary $\mathcal{N}=4$ super Yang-Mills theory to compute
the bi-fermion term. 
This term is evaluated by using Corrigan's
inner product formula and the result is
\begin{eqnarray}
\int \! d^4 x \ \frac{2}{\kappa} \mathrm{Tr}
\left[
\Lambda^{(0) \alpha A} \Lambda^{(0)}_{\alpha} {}^B
\right] m_{(AB)}
= - \frac{2\pi^2}{\kappa}
\mathrm{tr}_k
\left[
2 \bar{\mu}^A \mu^B + \mathcal{M}^{\prime \alpha A} \mathcal{M}'_{\alpha} {}^B
\right] m_{(AB)}.
\label{AS_field}
\end{eqnarray}
This completely agrees with the result of the string theory calculation (\ref{AS_string}) \cite{Ho}.
The instanton effective action for the (A,S)-deformed $\mathcal{N}=4$ super Yang-Mills theory is therefore
given by the sum of (\ref{aux_instanton}) and (\ref{AS_field}).
In a suitable basis of $SU(4)_R$, the mass matrix is diagonalized as
$m_{(AB)} = \mathrm{diag}(m_1, m_2, m_3, m_4)$.
If all the eigenvalues $m_1, \ldots, m_4$ are non-zero,
the deformed instanton effective action corresponds to
the one in massive deformation of
$\mathcal{N}=4$ super Yang-Mills theory.
In particular, if two eigenvalues are zero and the others take
the same non-zero value,
the deformed instanton effective action is that of
the mass deformed $\mathcal{N}=2^*$ super Yang-Mills theory
\cite{Ho, DoKhMa, BrFuMoTa}.
If one or three of the eigenvalues vanish and the others
are non-zero, the deformed instanton effective action is that of
the mass deformed $\mathcal{N}=4$ super Yang-Mills theory which preserves $\mathcal{N}=1$ supersymmetry \cite{DoHoKhMa}.

The instanton effective action for the (A,S)-deformed $\mathcal{N}=4$ super Yang-Mills theory is invariant under the
supersymmetry transformation (\ref{eq:n4susy3d}) 
with the following modifications for the transformations of $Y_{ma}, X_{\dot{\alpha}a}, \bar{X}_{\dot{\alpha}a}, \bar{\psi}^{\dot{\alpha}}_A$,
\begin{eqnarray}
\begin{aligned}
\delta' Y_{ma} =& 2 i \bar{\xi}^{\dot{\beta}} {}_C (\Sigma^a)^{AC}
(\sigma^m)^{\alpha} {}_{\dot{\beta}} \mathcal{M}'_{\alpha} {}^B m_{(AB)}, \\
\delta' X_{\dot{\alpha} a} =& 4 i \bar{\xi}_{\dot{\alpha} C}
(\Sigma^a)^{AC} \mu^B m_{(AB)}, \\
\delta' \overline{X}^{\dot{\alpha}} {}_a =& - 4 i
\bar{\xi}^{\dot{\alpha}} {}_C (\Sigma^a)^{AC} \bar{\mu}^B m_{(AB)}, \\
\delta' \bar{\psi}^{\dot{\alpha}}_A
=& 4 \bar{\xi}^{\dot{\alpha}}_C (\Sigma^a)^{BC} \chi_a m_{(AB)}
- 4 \bar{\xi}^{\dot{\alpha}}_C (\Sigma^a)^{BC} \phi_a^0 m_{(AB)}.
\end{aligned}
\end{eqnarray}
For these fields, the deformed supersymmetry transformation is given by $\delta + \delta'$ while for another fields,
the supersymmetry transformation is not modified.

Finally, when we introduce the (S,A) and (A,S)-backgrounds simultaneously, 
the following cross term would be induced in the low-energy effective Lagrangian
\begin{eqnarray}
\mathcal{L}_{\mathrm{cross}} \sim \frac{1}{\kappa} \mathrm{Tr} \left[ g^2 [\varphi^a, \varphi^b] C_{mnc} (\Sigma^a \bar{\Sigma}^b \Sigma^c)^{AB} m_{AB} \right],
\end{eqnarray}
where the overall coefficient is determined through the explicit calculations of the open string disk amplitudes.
This term is subleading order in $g$ both in the self-dual and anti-self-dual cases.
Therefore the instanton effective action for the self-dual case is simply given by the sum of (\ref{Inst_eff_D3}) and (\ref{AS_field}).

\section{Conclusions and discussion}
In this paper, we investigated the instanton effective action of the
$\mathcal{N} = 4$ super Yang-Mills theory deformed by the (S,A) and (A,S)-type
backgrounds
both from the string theory and field theory viewpoints.

In the string theory 
side, the instanton effective action is
interpreted as the effective action of the D$(-1)$-branes
in the D3-D$(-1)$ system.
This is evaluated from the disk amplitudes of open strings which have
at least one endpoint attached on the D$(-1)$-branes.
The R-R background is introduced by an insertion of 
the closed string vertex operators into the disk amplitudes.

In the case of (S,A)-type deformation,
the moduli $a'_m$ corresponding to the position of instantons get their 
mass at the second order of the background and hence these are fixed at the origin
in the D3-brane worldvolume. This implies that the translational invariance
is broken in the gauge theory.

On the other hand, from the viewpoint of the field theory,
the instanton effective action is obtained through the ADHM construction
of solutions in the (S,A)-deformed $\mathcal{N} = 4$ super Yang-Mills
theory. Since this theory has the translational invariance,
there is no mass term for $a'_m$ in the instanton effective action.
However, once we improve the spacetime action,
the instanton effective action
agrees with the string theory calculation.

We then compared the improved action for the (S,A)-deformed gauge
theory with the one derived from the gauge theory in the
$\Omega$-background.
The interaction terms contain explicit coordinate
dependence and are quite different from the improved (S,A)-deformed theory.
Nevertheless, 
we found that the equations of motion at the leading order
in the self-dual gauge field background
of the both theories coincide except for the one for
the anti-chiral fermions.
The instanton effective action in both theories also coincides.
This 
is similar to the $\mathcal{N} = 2$ case
\cite{ItNaSa2, proItNaSa}.

We also studied the effect of the (A,S)-type background.
It can be interpreted as the mass term for the chiral
fermion in the instanton effective action.
Choosing the eigenvalues of the mass matrix appropriately,
we obtained the instanton effective actions corresponding to massive deformations of $\mathcal{N}=4$ super Yang-Mills theory
with $\mathcal{N}=2$ or $\mathcal{N}=1$ supersymmetry.
Thus the R-R 3-form backgrounds provide various deformations of field theories and instantons.

In this paper, we also showed the supersymmetry invariance of the deformed
instanton effective action.
It is important to study the topological 
symmetry of the deformed action since the
BRST exactness of the instanton effective action is essential in the
calculation of the instanton partition function. 
This subject will be discussed in a separate paper.

Recently, it is recognized that the deformation 
by the R-R 3-form background also
plays an important role to investigate the heterotic-type I$'$ duality in 
study of eight-dimensional exotic instantons \cite{BiFeFrGaLePe, FuMoPo}.
It would be interesting to study general deformed gauge theories 
and their non-perturbative effects in various dimensions and R-R backgrounds.

\subsection*{Acknowledgments}
The work of K.~I. is supported in part by the Grant-in-Aid for Scientific
Research from Ministry of Education, Science,
Culture and Sports of Japan.
The work of H.~N. is supported by
the Postdoctoral Research Program of Sungkyunkwan University (2009)
and is the result of research activities (Astrophysical Research
Center for the Structure and Evolution of the Cosmos (ARCSEC))
supported by KOSEF.
The work of T.~S. is supported by the Global
Center of Excellence Program by MEXT, Japan through the
"Nanoscience and Quantum Physics" Project of the Tokyo
Institute of Technology, and by Iwanami Fujukai Foundation.
The work of S.~S. is supported by the Japan Society for the Promotion of Science (JSPS) Research Fellowship.

\begin{appendix}
\section{Detailed calculation of the disk amplitudes}
In this appendix, we present the detailed calculations of the open string
disk amplitudes
including an insertion of the vertex operator corresponding to the closed string backgrounds.
Our conventions and notations on the type IIB string
worldsheet variables are found in \cite{ItSa, ItKoSa, ItNaSa1}.
We make use of the NSR formalism to calculate the amplitudes
in ten-dimensional Euclidean spacetime.
We are interested
in open string disk amplitudes with at least one edge of the disk is
attached on the D$(-1)$-branes.
The open string vertex operators are inserted at the boundary of the disk
parametrized by real coordinates
$y_i$ while the closed string vertex operator is inside the disk
parametrized by a complex coordinate $z$.
In general, the $(n+2)$-point disk amplitude with $n$ open string vertex
operators $V_{\Phi_i}^{(q_i)} (y_i)$ and one closed string vertex operator
$V^{(-1/2,-1/2)}_{\mathcal{F}} (z,\bar{z})$ with at least one boundary on the D$(-1)$-branes
is given by
\begin{eqnarray}
\langle \! \langle
V_{\Phi_1}^{(q_1)} \cdots V_{\Phi_n}^{(q_n)} V_{\mathcal{F}}^{(-1/2,-1/2)}
\rangle \! \rangle
= C_{-1}
\int \! \frac{\prod^n_{i=1} d y_i d z d \bar{z}}{d V_{\mathrm{CKG}}}
\langle
V_{\Phi_1}^{(q_1)} (y_1) \cdots V_{\Phi_n}^{(q_n)} (y_n)
V_{\mathcal{F}}^{(-1/2,-1/2)} (z, \bar{z})
\rangle,
\nonumber \\
\end{eqnarray}
where $C_{-1} = \frac{1}{2 \pi^2
\alpha^{\prime 2}} \frac{1}{\kappa g^2_0}$ is the normalization factor of the disk amplitudes.
$d V_{\mathrm{CKG}}$ is an $SL(2,\mathbf{R})$ invariant volume factor of the conformal Killing group to
fix three positions of the vertex operators.
We fix $z \to i, \bar{z} \to - i$ and one of the $y_i$ to
$\infty$ in the following calculations.
Note that because we are considering disk amplitudes, all the
$\phi$-charges in the bosonic ghost should add up to $-2$.
In the following, we separately calculate the disk amplitudes with an insertion of the (S,A) and
(A,S)-background.

\subsection{(S,A)-background}
As we mentioned in section 2,
there are only two
non-zero amplitudes containing one vertex operator 
of the (S,A)-background, which are given as
\begin{eqnarray}
\langle \! \langle V^{(-1/2)}_{\mathcal{M}'}
V^{(-1/2)}_{\mathcal{M}'} V^{(-1/2,-1/2)}_{\mathcal{F}}
\rangle \! \rangle, \quad
\langle \! \langle
V^{(0)}_Y V^{(0)}_{a'} V^{(-1/2,-1/2)}_{\mathcal{F}}
\rangle \! \rangle.
\label{SA_amplitudes2}
\end{eqnarray}
The first amplitude is 
\begin{eqnarray}
\langle \! \langle V^{(-1/2)}_{\mathcal{M}'}
V^{(-1/2)}_{\mathcal{M}'} V^{(-1/2,-1/2)}_{\mathcal{F}}
\rangle \! \rangle
&=&
\frac{1}{2\pi^2 \alpha^{\prime 2}} \frac{1}{\kappa g_0^2}
(2\pi \alpha')^2 g_0^2
(\pi^2)
\mathrm{tr}_k
\left[
\mathcal{M}^{\prime \alpha A} \mathcal{M}^{\prime \beta B}
(2\pi \alpha')^{\frac{1}{2}} \mathcal{F}^{(\gamma \delta) [CD]}
\right]
\nonumber \\
& & \times \int^{y_1}_{- \infty} \! d y_2 \ (y_1 - z) (y_1 - \bar{z}) (z
- \bar{z})
\langle e^{- \frac{1}{2} \phi (y_1)} e^{- \frac{1}{2} \phi (y_2)}
e^{- \frac{1}{2} \phi (z)} e^{- \frac{1}{2} \phi (\bar{z})}
\rangle
\nonumber \\
& & \times
\langle S_{\alpha} (y_1) S_{\beta} (y_2) S_{\gamma} (z) S_{\delta} (\bar{z})
\rangle
\langle S_A (y_1) S_B (y_2) S_C (z) S_D (\bar{z})
\rangle.
\end{eqnarray}
Here the correlators are calculated as
\begin{eqnarray}
\langle e^{- \frac{1}{2} \phi (y_1)} e^{- \frac{1}{2} \phi (y_2)}
e^{- \frac{1}{2} \phi (z)} e^{- \frac{1}{2} \phi (\bar{z})}
\rangle
\label{ghost_correlator}
&=&
\left[
(y_1 - y_2) (y_1 - z) (y_1 - \bar{z}) (y_2 - z) (y_2 - \bar{z}) (z - \bar{z})
\right]^{-\frac{1}{4}}, \\
\langle S_{\alpha} (y_1) S_{\beta} (y_2) S_{\gamma} (z) S_{\delta} (\bar{z})
\rangle
&=& \left[
(y_1 - y_2) (y_1 - z) (y_1 - \bar{z}) (y_2 - z) (y_2 - \bar{z}) (z - \bar{z})
\right]^{-\frac{1}{2}}
\nonumber \\
& & 
\times
\left\{
\varepsilon_{\alpha \delta} \varepsilon_{\beta \gamma}
(y_1 - y_2) (z - \bar{z}) - \varepsilon_{\alpha \beta} \varepsilon_{\gamma \delta}
(y_1 - \bar{z}) (y_2 - z)
\right\}, \\
\langle S_A (y_1) S_B (y_2) S_C (z) S_D (\bar{z})
\rangle
&=& \varepsilon_{ABCD}
\left[
(y_1 - y_2) (y_1 - z) (y_1 - \bar{z}) (y_2 - z) (y_2 - \bar{z}) (z - \bar{z})
\right]^{-\frac{1}{4}}.
\nonumber \\
\end{eqnarray}
We have omitted the overall cocycle factor  \cite{KoLeLeSaWa}
which will be multiplied at the end of calculations.
After fixing the positions of vertex operators such as
$y_1 \to \infty, z \to i, \bar{z} \to - i$, and
integrating over $y_2$, we get
\begin{eqnarray}
\langle \! \langle V^{(-1/2)}_{\mathcal{M}'}
V^{(-1/2)}_{\mathcal{M}'} V^{(-1/2,-1/2)}_{\mathcal{F}}
\rangle \! \rangle
= \frac{2\pi^2}{\kappa} \mathrm{tr}_k
\left[
\frac{1}{2}
(\overline{\Sigma}^a)_{AB}
\mathcal{M}^{\prime A}_{\alpha}
\mathcal{M}^{\prime B}_{\beta} (2 \pi i^2) (2 \pi \alpha')^{\frac{1}{2}}
(\overline{\Sigma}^a)_{CD}
\mathcal{F}^{(\alpha \beta) [CD]}
\right],
\nonumber \\
\end{eqnarray}
where we have used the relation $ \varepsilon_{ABCD} = \frac{1}{2}
(\overline{\Sigma}^a)_{AB} (\overline{\Sigma}^a)_{CD}$
and multiplied the cocycle factor which has been evaluated as $+i$.

The second amplitude in (\ref{SA_amplitudes2}) is
\begin{eqnarray}
\langle \! \langle
V^{(0)}_Y V^{(-1)}_{a'} V^{(-1/2,-1/2)}_{\mathcal{F}}
\rangle \! \rangle
&=&
\frac{1}{2\pi^2 \alpha^{\prime 2}} \frac{1}{\kappa g_0^2}
(2\pi \alpha')^2 g_0^2
\left(
\frac{4\pi^2}{\sqrt{2}}
\right)
\mathrm{tr}_k
\left[
Y_{ma} a'_n (2 \pi \alpha')^{\frac{1}{2}} \mathcal{F}^{(\alpha \beta)[AB]}
\right]
\nonumber \\
& & \times \int^{y_1}_{- \infty} \! d y_2 \ (y_1 - z) (y_1 - \bar{z}) (z
- \bar{z})
\langle e^{- \phi (y_2)} e^{- \frac{1}{2} \phi (z)} e^{- \frac{1}{2}
\phi (\bar{z})} \rangle
\nonumber \\
& & \times
\langle
\psi^m \psi^a (y_1)
\psi^n (y_2) S_{\alpha} (z) S_A (z) S_{\beta} (\bar{z}) S_B (\bar{z})
\rangle.
\end{eqnarray}
The correlator of the ghost fields is evaluated as
\begin{eqnarray}
\langle e^{- \phi (y_2)} e^{- \frac{1}{2} \phi (z)} e^{- \frac{1}{2}
\phi (\bar{z})} \rangle
= (y_1 - z)^{- \frac{1}{2}} (y_1 - \bar{z})^{- \frac{1}{2}} (z -
\bar{z})^{- \frac{1}{4}}.
\end{eqnarray}
The correlator including spin fields is calculated as
\begin{eqnarray}
& & \langle
\psi^m \psi^a (y_1)
\psi^n (y_2) S_{\alpha} (z) S_A (z) S_{\beta} (\bar{z}) S_B (\bar{z})
\rangle \nonumber \\
& & = \frac{\delta^{mn}}{(y_1 - y_2)}
\langle
\psi^a (y_2) S_{\alpha} (z) S_A (z) S_{\beta} (\bar{z}) S_B (\bar{z})
\rangle \nonumber \\
& &
+ \frac{1}{2}
(y_1 - z) (\sigma^m)_{\alpha \dot{\alpha}} (\overline{\Sigma}^a)_{AC}
\langle
\psi^n (y_2) S^{\dot{\alpha}} (z) S^C (z) S_{\beta} (\bar{z}) S_B (\bar{z})
\rangle
\nonumber \\
& &
+ \frac{1}{2}
(y_1 - \bar{z}) (\sigma^m)_{\beta \dot{\beta}}
(\overline{\Sigma}^a)_{BD}
\langle
\psi^n (y_2) S_{\alpha} (z) S_A (z) S^{\dot{\beta}} (\bar{z}) S^D (\bar{z})
\rangle,
\end{eqnarray}
where we have used the fact that $\psi^m \psi^a$ acts on the other fields in the correlator as the ten-dimensional
Lorentz generator \cite{KoLeLeSaWa}.
The first term in the above equation is proportional to $\varepsilon_{\alpha \beta}$ and will vanish when it is contracted
with $\mathcal{F}^{(\alpha \beta)}$ while the second and the third terms are evaluated as
\begin{eqnarray}
& &
\frac{1}{2 \sqrt{2}} (y_1 - z)^{-1}
\left[
(y_2 - z)^{-\frac{1}{2}} (y_2 - \bar{z})^{- \frac{1}{2}}
(z - \bar{z})^{- \frac{3}{4}}
\right]
(\sigma^m)_{\alpha \dot{\alpha}} (\bar{\sigma}^n)^{\dot{\alpha}}
{}_{\beta} (\overline{\Sigma}^a)_{AB}
\nonumber \\
& & +
\frac{1}{2 \sqrt{2}} (y_1 - \bar{z})^{-1}
\left[
(y_2 - z)^{-\frac{1}{2}} (y_2 - \bar{z})^{- \frac{1}{2}}
(z - \bar{z})^{- \frac{3}{4}}
\right]
(\sigma^m)_{\beta \dot{\beta}} (\bar{\sigma}^n)^{\dot{\beta}}
{}_{\alpha} (\overline{\Sigma}^a)_{BA}.
\end{eqnarray}
Here we have used the following relations
\begin{eqnarray}
\langle \psi^n (y_2) S^{\dot{\alpha}} (z) S_{\beta} (\bar{z})
\rangle
&=& \frac{1}{\sqrt{2}} (\bar{\sigma}^n)^{\dot{\alpha}} {}_{\beta}
(y_2 - z)^{- \frac{1}{2}} (y_2 - \bar{z})^{- \frac{1}{2}},
\\
\langle \psi^n (y_2) S_{\alpha} (z) S^{\dot{\beta}} (\bar{z})
\rangle
&=& \frac{1}{\sqrt{2}} (\sigma^n)_{\alpha} {}^{\dot{\beta}}
(y_2 - z)^{-\frac{1}{2}} (y_2 - \bar{z})^{- \frac{1}{2}},
\\
\langle S_A (z) S^B (\bar{z}) \rangle
&=& \delta_A {}^B (z - \bar{z})^{- \frac{3}{4}}.
\end{eqnarray}
After evaluating
the $y_2$ integration,
the second amplitude in (\ref{SA_amplitudes2}) is obtained as
\begin{eqnarray}
\!\!\!\!\!\!\!\!\!\!  \langle \! \langle
V^{(0)}_Y V^{(-1)}_{a'} V^{(-1/2,-1/2)}_{\mathcal{F}}
\rangle \! \rangle
\!=\! \frac{2\pi^2}{\kappa}
\mathrm{tr}_k \!\!
\left[
- \frac{i}{\sqrt{2}} (2 \pi i)
(\sigma^{mn})_{\alpha \beta} (\overline{\Sigma}^a)_{AB}
Y_{ma} a'_n (2 \pi \alpha')^{\frac{1}{2}} \mathcal{F}^{(\alpha \beta) [AB]}
\right],
\end{eqnarray}
where we have multiplied the cocycle factor which has been calculated as $+i$.

\subsection{(A,S)-background}
The non-zero amplitudes including one (A,S)-background vertex operator are
\begin{eqnarray}
\langle \! \langle
V^{(-1/2)}_{\mathcal{M}'}
V^{(-1/2)}_{\mathcal{M}'}
V^{(-1/2,-1/2)}_{\mathcal{F}}
\rangle \! \rangle,
\qquad
\langle \! \langle
V^{(-1/2)}_{\bar{\mu}}
V^{(-1/2)}_{\mu}
V^{(-1/2,-1/2)}_{\mathcal{F}}
\rangle \! \rangle.
\end{eqnarray}
The first amplitude is
\begin{eqnarray}
\langle \! \langle
V^{(-1/2)}_{\mathcal{M}'} V^{(-1/2)}_{\mathcal{M}'}
V^{(-1/2,-1/2)}_{\mathcal{F}}
\rangle \! \rangle
&=& \frac{1}{2\pi^2 \alpha'^2}
\frac{1}{\kappa g^2_0} (2\pi \alpha')^2 \pi^2 g_0^2
\mathrm{tr}_k
\left[
\mathcal{M}^{\prime \alpha A} \mathcal{M}^{\prime \beta B}
(2\pi \alpha')^{\frac{1}{2}} \mathcal{F}^{[\dot{\alpha} \dot{\beta}]} {}_{(CD)}
\right] \nonumber \\
& & \!\!\! \times \int^{y_1}_{- \infty} \! d y_2 \
(y_1 - z) (y_1 - \bar{z}) (z - \bar{z})
\langle
e^{- \frac{1}{2} \phi (y_1)}
e^{- \frac{1}{2} \phi (y_2)}
e^{- \frac{1}{2} \phi (z)}
e^{- \frac{1}{2} \phi (\bar{z})}
\rangle
\nonumber \\
& &\!\!\! \times \langle S_{\alpha} (y_1) S_{\beta} (y_2) \rangle
\langle S_{\dot{\alpha}} (z) S_{\dot{\beta}} (\bar{z}) \rangle
\langle
S_A (y_1) S_B (y_2) S^C (z) S^D (\bar{z})
\rangle.
\end{eqnarray}
The spin field correlators in the above equation are evaluated as
\begin{eqnarray}
\langle S_{\alpha} (y_1) S_{\beta} (y_2) \rangle &=& \varepsilon_{\alpha
\beta} (y_1 - y_2)^{- \frac{1}{2}}, \\
\langle S_{\dot{\alpha}} (z) S_{\dot{\beta}} (\bar{z}) \rangle &=&
\varepsilon_{\dot{\alpha} \dot{\beta}} (z - \bar{z})^{-\frac{1}{2}},
\label{spin_correlator1}
\\
\langle
S_A (y_1) S_B (y_2) S^C (z) S^D (\bar{z})
\rangle
&=&
\left[
(y_1 - y_2) (y_1 - z) (y_1 - \bar{z})
(y_2 - z) (y_2 - \bar{z})
\right]^{-\frac{3}{4}}
(z - \bar{z})^{- \frac{1}{4}}
\nonumber \\
& & \times
\left[
- (y_1 - \bar{z}) (y_2 - z) \delta_A {}^C \delta_B {}^D
+ (y_1 - z) (y_2 - \bar{z}) \delta_A {}^D \delta_B {}^C
\right].
\label{spin_correlator2}
\nonumber \\
\end{eqnarray}
The correlator of the ghost part is given in (\ref{ghost_correlator}).
After performing
the $y_2$ integration
, the result is
\begin{eqnarray}
\langle \! \langle
V^{(-1/2)}_{\mathcal{M}'} V^{(-1/2)}_{\mathcal{M}'}
V^{(-1/2,-1/2)}_{\mathcal{F}}
\rangle \! \rangle
= \frac{2\pi^2}{\kappa} \mathrm{tr}_k
\left[
2 \mathcal{M}^{\prime \alpha A} \mathcal{M}'_{\alpha} {}^B
\pi i (2 \pi \alpha')^{\frac{1}{2}} \mathcal{F}^{[\dot{\alpha} \dot{\beta}]} {}_{(AB)} \varepsilon_{\dot{\alpha} \dot{\beta}}
\right].
\end{eqnarray}
Here the cocycle factor has been evaluated as $+1$.

The second amplitude is
\begin{eqnarray}
\langle \! \langle
V^{(-1/2)}_{\bar{\mu}} V^{(-1/2)}_{\mu}
V^{(-1/2,-1/2)}_{\mathcal{F}}
\rangle \! \rangle
&=& \frac{1}{2\pi^2 \alpha'^2}
\frac{1}{\kappa g^2_0}
(2\pi \alpha')^2 g_0^2 \pi^2
\left(
\frac{2}{\sqrt{2}^2}
\right)
\mathrm{tr}_k
\left[
\bar{\mu}^A \mu^B (2 \pi \alpha')^{\frac{1}{2}}
\mathcal{F}^{[\dot{\alpha} \dot{\beta}]} {}_{(CD)}
\right] \nonumber \\
& & \!\!\!\! \times \int^{y_1}_{- \infty}
\! d y_2 \ (y_1 - z) (y_2 - \bar{z}) (z - \bar{z})
\langle
e^{- \frac{1}{2} \phi (y_1)}
e^{- \frac{1}{2} \phi (y_2)}
e^{- \frac{1}{2} \phi (z)}
e^{- \frac{1}{2} \phi (\bar{z})}
\rangle
\nonumber \\
& & \!\!\!\! \times
\langle
\overline{\Delta} (y_1) \Delta (y_2)
\rangle
\langle
S_{\dot{\alpha}} (z) S_{\dot{\beta}} (\bar{z})
\rangle
\langle
S_A (y_1) S_B (y_2) S^C (z) S^D (\bar{z})
\rangle.
\end{eqnarray}
The correlators of the ghost and spin fields have been evaluated in (\ref{ghost_correlator}), (\ref{spin_correlator1}), (\ref{spin_correlator2}).
The correlator of the twist field is given as
\begin{eqnarray}
\langle \bar{\Delta} (y_1) \Delta (y_2) \rangle
= (y_1 - y_2)^{- \frac{1}{2}}.
\end{eqnarray}
After performing the 
$y_2$ integration, we find
\begin{eqnarray}
\langle \! \langle
V^{(-1/2)}_{\bar{\mu}} V^{(-1/2)}_{\mu}
V^{(-1/2,-1/2)}_{\mathcal{F}}
\rangle \! \rangle
=
\frac{2\pi^2}{\kappa}
\mathrm{tr}_k
\left[
2 \bar{\mu}^A \mu^B
\pi i (2 \pi \alpha')^{\frac{1}{2}} \mathcal{F}^{[\dot{\alpha} \dot{\beta}]} {}_{(AB)} \varepsilon_{\dot{\alpha} \dot{\beta}}
\right].
\end{eqnarray}
Here the cocycle factor has been absorbed into the redefinition of $\mu$ and $\bar{\mu}$.

\section{The ADHM construction}
Here we briefly summarize the ADHM construction of instantons with topological number $k$, where $k$ is a positive integer. We introduce
the $(N+2k)\times 2k$ matrix $\Delta_{\lambda j\dot{\alpha}}$
which is given by
\begin{equation}
\Delta_{\lambda j\dot{\alpha}}
=a_{\lambda j\dot{\alpha}}+b_{\lambda j}{}^{\beta}\sigma_{m\beta\dot{\alpha}}x^{m},
\end{equation}
where $\alpha, \dot{\alpha} = 1,2$, $\lambda=1,2,\ldots,N+2k$ and $i,j=1,2,\ldots,k$.
$a_{\lambda j\dot{\alpha}}$ and $b_{\lambda j}{}^{\beta}$ are
the constant matrices. They are decomposed as
\begin{equation}
a_{\lambda j\dot{\alpha}}=\binom{w_{uj\dot{\alpha}}}
{(a'_{\alpha\dot{\alpha}})_{ij}}, \quad
b_{\lambda j}{}^{\beta}=\binom{0}{\delta_{ij}\delta_{\alpha}{}^{\beta}}, \quad
\lambda=u+i\alpha,\quad u=1,2,\ldots,N.
\end{equation}
These matrices $w,a'$ are called bosonic ADHM moduli.
$\Delta$ obeys the condition
\begin{equation}
\bar{\Delta}^{\dot{\alpha}\lambda}_{i}\Delta_{\lambda j\dot{\beta}}
=(f^{-1})_{ij}\delta^{\dot{\alpha}}{}_{\dot{\beta}},\quad
f_{ij}=\biggl[\frac{1}{2}\bar{w}^{\dot{\alpha}u}_i w_{uj\dot{\alpha}}
+(x_{m}\delta_{ik}+(a'_{m})_{ik})(x^{m} \delta_{kj}+(a'^{m})_{kj}) \biggr]^{-1},
\label{ADHM}
\end{equation}
where the barred matrix denotes its Hermitian conjugate.
The first equation in (\ref{ADHM}) is called ADHM constraint.
In terms of the ADHM moduli 
$a'$, $w$, and $\bar{w}$, 
the ADHM constraint
can be rewritten as
\begin{equation}
(\tau^{c})^{\dot{\alpha}}_{~\dot{\beta}}
(\bar{w}^{\dot{\beta}}w_{\dot{\alpha}}
+\bar{a}^{\prime\dot{\beta}\alpha}a'_{\alpha\dot{\alpha}})=0,\quad
a'_{m}=\bar{a}'_{m}. \label{ADHM2}
\end{equation}
We also introduce $(N+2k)\times N$ matrix $U$ which satisfies
\begin{equation}
\bar{\Delta}U=0,\quad \bar{U}U=\boldsymbol{1}_{N},\quad
U\bar{U}+\Delta_{\dot{\alpha}} f\bar{\Delta}^{\dot{\alpha}}=
\boldsymbol{1}_{N+2k},
\end{equation}
where $\boldsymbol{1}_{N}$ is the $N \times N$ identity matrix.
Then the self-dual equation $F_{mn}^{(0)(-)}=0$ is solved in terms of $U$ as
\begin{equation}
A_{m}^{(0)}=-i\bar{U}\partial_{m}U.
\end{equation}
The corresponding gauge field strength $F_{mn}^{(0)}$ is given by
\begin{equation}
F_{mn}^{(0)}=-4i\bar{U}b^{\alpha}(\sigma_{mn})_{\alpha}{}^{\beta}f
\bar{b}_{\beta}U.
\label{SDA}
\end{equation}

We discuss the fermionic part in $\mathcal{N}=4$ theory.
We solve the Dirac equation
$\bar{\sigma}^{m}\nabla_{m}\Lambda^{(0)A}=0$,
where $\nabla_{m}$ denotes the covariant derivative in the self-dual instanton background.
The ansatz of the solution is
\begin{equation}
\Lambda^{(0)A}_{\alpha}=\bar{U}
(\mathcal{M}^{A}f\bar{b}_{\alpha}-b_{\alpha}f\bar{\mathcal{M}}^{A})U,
\label{af}
\end{equation}
where $\mathcal{M}^{A}$ is the $(N+2k)\times k$ constant matrix.
Plugging \eqref{af} into the Dirac equation, we obtain
\begin{equation}
\bar{\sigma}^{m}\nabla_{m}\Lambda^{(0)A}=2\bar{U}b^{\alpha}f
(\bar{\mathcal{M}}^{A}\Delta+\bar{\Delta}\mathcal{M}^{A})f\bar{b}_{\alpha}U.
\end{equation}
Then we have the fermionic ADHM constraint: 
\begin{equation}
\bar{\mathcal{M}}^{A}\Delta+\bar{\Delta}\mathcal{M}^{A}=0,
\label{fADHM}
\end{equation}
or equivalently
\begin{equation}
\bar{\mu}^{A}w_{\dot{\alpha}}+\bar{w}_{\dot{\alpha}}\mu^{A}
+[\mathcal{M}^{\prime\alpha A},a'_{\alpha\dot{\alpha}}]=0,
\quad \mathcal{M}^{\prime A}=\bar{\mathcal{M}}^{\prime A},
\label{fADHM2}
\end{equation}
where 
we have decomposed $\mathcal{M}^{A}$ as
\begin{equation}
\mathcal{M}^{A}_{\lambda j}
=\binom{\mu^{A}_{uj}}{(\mathcal{M}^{\prime A}_{\alpha})_{ij}},
\end{equation}
$\mu^{A}_{uj}$ and $\mathcal{M}^{\prime A}_{\alpha}$ are called the fermionic ADHM moduli.

Now we solve the equation of motion for the scalar field $\varphi_{a}^{(0)}$ (\ref{scalar_eq}) with the asymptotic boundary condition
$\lim_{|x|\to\infty}\varphi_{a}^{(0)}=\phi^0_{a}$.
First we consider the case of $C=0$.
The ansatz of the solution is
\begin{equation}
\varphi_{a}^{(0)}=-\frac{1}{4}(\bar{\Sigma}^{a})_{AB}\bar{U}\mathcal{M}^{A}f
\bar{\mathcal{M}}^{B}U
+\bar{U}\begin{pmatrix} \phi^0_{a} & 0 \\
0 & \chi_{a}\boldsymbol{1}_{2}
\end{pmatrix}U.
\end{equation}
Computing $\nabla^{2}\varphi_{a}^{(0)}$ explicitly,
we obtain
\begin{align}
\nabla^{2}\varphi_{a}^{(0)}
&=
(\bar{\Sigma}^{a})_{AB}\Lambda^{(0)A}\Lambda^{(0)B}
+4\bar{U}bf\Biggl[\frac{1}{4}(\bar{\Sigma}^{a})_{AB}
\bar{\mathcal{M}}^{A}\mathcal{M}^{B}
-\{f^{-1},\chi_{a}\}+\bar{\Delta}^{\dot{\alpha}}
\begin{pmatrix} \phi^0_{a} & 0 \\
0 & \chi_{a}\boldsymbol{1}_{2}
\end{pmatrix}
\Delta_{\dot{\alpha}}\Biggr]f\bar{b}U
\notag\\
&=
(\bar{\Sigma}^{a})_{AB}\Lambda^{(0)A}\Lambda^{(0)B}
+4\bar{U}bf\biggl(\frac{1}{4}(\bar{\Sigma}^{a})_{AB}
\bar{\mathcal{M}}^{A}\mathcal{M}^{B}-\mathbf{L}\chi_{a}
+\bar{w}^{\dot{\alpha}}\phi^0_{a}w_{\dot{\alpha}}\biggr)f\bar{b}U,
\end{align}
where $\mathbf{L}\chi_{a}$ is defined by
\begin{equation}
\mathbf{L}\chi_{a}\equiv
\frac{1}{2}\bigl\{\bar{w}^{\dot{\alpha}}w_{\dot{\alpha}},\chi_{a}\bigr\}
+\Bigl[a'_{m},[a^{\prime m},\chi_{a}]\Bigr].
\label{definition_bfL}
\end{equation}
Then $\varphi_{a}^{(0)}$ satisfies the equation of motion if $\chi_{a}$ satisfies
\begin{equation}
\mathbf{L}\chi_{a}=\frac{1}{4}(\bar{\Sigma}^{a})_{AB}
\bar{\mathcal{M}}^{A}\mathcal{M}^{B}
+\bar{w}^{\dot{\alpha}}\phi^0_{a} w_{\dot{\alpha}}.
\end{equation}
In the case of $C\neq 0$ we change the ansatz of the solution as
\begin{equation}
\varphi_{a}^{(0)}=-\frac{1}{4}(\bar{\Sigma}^{a})_{AB}\bar{U}\mathcal{M}^{A}f
\bar{\mathcal{M}}^{B}U
+\bar{U}\begin{pmatrix} \phi^0_{a} & 0 \\
0 & \chi_{a}\boldsymbol{1}_{2}+\boldsymbol{1}_{k}C_{a}
\end{pmatrix}U,
\end{equation}
where $C_{a}$ is the $2\times 2$ matrix of which components are
$(C_{a})_{\alpha}^{~\beta}= (\sigma^{m n})_{\alpha} {}^{\beta} C_{m n a}$. Now one can show that
\begin{equation}
\begin{split}\label{eom1}
\nabla^{2}\varphi_{a}^{(0)}
&=
(\bar{\Sigma}^{a})_{AB}\Lambda^{(0)A}\Lambda^{(0)B}
+4\bar{U}bf\Biggl[\frac{1}{4}(\bar{\Sigma}^{a})_{AB}
\bar{\mathcal{M}}^{A}\mathcal{M}^{B}\\
&\qquad
{}-2f^{-1}C_{a}-\{f^{-1},\chi_{a}\}+\bar{\Delta}^{\dot{\alpha}}
\begin{pmatrix} \phi^0_{a} & 0 \\
0 & \chi_{a}\boldsymbol{1}_{2}+\boldsymbol{1}_{k}C_{a}
\end{pmatrix}
\Delta_{\dot{\alpha}}\Biggr]f\bar{b}U.
\end{split}
\end{equation}
The second term in the square brackets in \eqref{eom1} becomes the deformation term $-iC^{mna}F_{mn}^{(0)}$ in
the equation of motion by using \eqref{SDA}.
The $C$-dependent part in the last term in the square brackets becomes
\begin{equation}
\bar{\Delta}^{\dot{\alpha}}
\begin{pmatrix} 0 & 0 \\
0 & \boldsymbol{1}_{k}C_{a}
\end{pmatrix}
\Delta_{\dot{\alpha}}
=(\bar{a}'+x^{m}\bar{\sigma}_{m})^{\dot{\alpha}\alpha}(C_{a})_{\alpha}{}^{\beta}
(a'+x^{n}\sigma_{n})_{\beta\dot{\alpha}}
=C^{mn a}[a'_{m},a'_{n}].
\end{equation}
Then we obtain
\begin{equation}
\begin{split}
\nabla^{2}\varphi_{a}^{(0)}
&=
(\bar{\Sigma}^{a})_{AB}\Lambda^{(0)A}\Lambda^{(0)B}-iC^{mn a}F_{mn}^{(0)}\\
&\qquad
+4\bar{U}bf\biggl(\frac{1}{4}(\bar{\Sigma}^{a})_{AB}
\bar{\mathcal{M}}^{A}\mathcal{M}^{B}
-\mathbf{L}\chi_{a}
+\bar{w}^{\dot{\alpha}}\phi^0_{a} w_{\dot{\alpha}}
+C^{mn a}[a'_{m},a'_{n}]
\biggr)f\bar{b}U.
\end{split}
\end{equation}
Hence $\varphi_{a}^{(0)}$ is the solution of the deformed equation of motion
if $\chi_{a}$ satisfies
\begin{equation}
\mathbf{L}\chi_{a}=\frac{1}{4}(\bar{\Sigma}^{a})_{AB}
\bar{\mathcal{M}}^{A}\mathcal{M}^{B}
+\bar{w}^{\dot{\alpha}}\phi^0_{a} w_{\dot{\alpha}}
+C^{mn a}[a'_{m},a'_{n}].
\end{equation}

\end{appendix}

\end{document}